\documentclass[journal, twocolumn, oneside, a4paper]{IEEEtran}

\usepackage{cite}

\ifCLASSINFOpdf
  \usepackage[pdftex]{graphicx}
   \graphicspath{{..1/pdf/}{../jpeg/}}
   \DeclareGraphicsExtensions{.pdf,.jpeg,.png}
\else
   \usepackage[dvips]{graphicx}
   \graphicspath{{../eps/}}
   \DeclareGraphicsExtensions{.eps}
\fi

\usepackage{cite}
\usepackage{comment} 
\usepackage[cmex10]{amsmath}
\usepackage{amsfonts, amsmath, amsthm, amstext, amssymb, mathrsfs, graphicx, color, url}
\usepackage[colorlinks=false, linkcolor=blue]{hyperref}
\usepackage{array}
\usepackage{mdwmath}
\usepackage{mdwtab}
\usepackage{algpseudocode}
\usepackage{algorithm}
\usepackage{url}
\usepackage{setspace}
\usepackage{footnote}
\usepackage[skip=0pt]{caption}
\usepackage{float}
\usepackage{epstopdf}

\def\IC{{\mathbb C}}

\def\IS{{\mathbb S}}

\def\calI{{\mathcal I}}

\def\calK{{\mathcal K}}
\def\calL{{\mathcal L}}

\def\calO{{\mathcal O}}

\def\calU{{\mathcal U}}

\def\bA{{\pmb A}}
\def\bB{{\pmb B}}

\def\bH{{\pmb H}}
\def\bI{{\pmb I}}

\def\bK{{\pmb K}}
\def\bL{{\pmb L}}
\def\bM{{\pmb M}}

\def\bP{{\pmb P}}
\def\bQ{{\pmb Q}}
\def\bR{{\pmb R}}
\def\bS{{\pmb S}}
\def\bT{{\pmb T}}
\def\bU{{\pmb U}}
\def\bV{{\pmb V}}
\def\bW{{\pmb W}}
\def\bX{{\pmb X}}

\def\bZ{{\pmb Z}}

\def\bf{{\pmb f}}

\def\by{{\pmb y}}

\newcommand{\bSig}{  \pmb{\Sigma}  }

\newcommand{\bPsi}{  \pmb{\Psi}  }

\newcommand{\bLam}{  \pmb{\Lambda}  }
\newcommand{\bTh}{  \pmb{\Theta}  }

\newcommand{\upperRomannumeral}[1]{\uppercase\expandafter{\romannumeral#1}}

\newcommand{\st}{ \textup{s. t.} }
\newcommand{\argmin}{ \textup{argmin} }
\newcommand{\argmax}{ \textup{argmax} }

\def\tr{{\textup{tr}}}

\newcommand{\lrb}[1]{ \lbrace #1 \rbrace }

\newtheorem{lemma}{Lemma}

\newtheorem{proposition}{Proposition}

\newtheorem{corollary}{Corollary}

\newtheorem{definition}{Definition}

\newtheorem{remark}{Remark}

\newtheorem{assumption}{Assumption}

\begin{document}

\title{Sum-rate Maximization in Sub-$28$ GHz Millimeter-Wave MIMO Interfering Networks}

\author{Hadi~Ghauch,~\IEEEmembership{Student Member,~IEEE,}
        Taejoon~Kim,~\IEEEmembership{Member,~IEEE,} \\
        Mats~Bengtsson,~\IEEEmembership{Senior Member,~IEEE,}
        and~Mikael~Skoglund,~\IEEEmembership{Senior Member,~IEEE} 

\thanks{H. Ghauch, M. Bengtsson and M. Skoglund are with the School of Electrical Engineering and the ACCESS Linnaeus Center, KTH Royal Institute of Technology, Stockholm, Sweden. E-mails: ghauch@kth.se, mats.bengtsson@ee.kth.se, skoglund@kth.se }
\thanks{T. Kim is with the State Key Laboratory of Millimeter Wave and the Department of Electronic Engineering, City University of Hong Kong, Kowloon, Hong Kong (e-mail: taejokim@cityu.edu.hk) } }

\markboth{}{Ghauch \MakeLowercase{\textit{et al.}}: Sum-rate Maximization in Sub-$28$ GHz Millimeter-Wave MIMO Interfering Networks}

\maketitle

\begin{abstract}
MIMO systems in the lower part of the millimetre-wave spectrum band (i.e., below $28$ GHz) do not exhibit enough directivity and selectively, as their counterparts in higher bands of the spectrum (i.e., above $60$ GHz), and thus still suffer from the detrimental effect of interference, on the system sum-rate.  As such systems exhibit large numbers of antennas and short coherence times for the channel, traditional methods of distributed coordination are ill-suited, and the resulting communication overhead would offset the gains of coordination.
In this work, we propose algorithms for tackling the sum-rate maximization problem, that are designed to address the above limitations.  
We derive a lower bound on the sum-rate, a so-called DLT bound  (i.e., a difference of log and trace), shed light on its tightness, and highlight its decoupled nature at both the transmitters and receivers. Moreover, we derive the solution to each of the subproblems, that we dub non-homogeneous waterfilling (a variation on the MIMO waterfilling solution), and underline an inherent desirable feature: its ability to turn-off streams exhibiting low-SINR, and contribute to greatly speeding up the convergence of the proposed algorithm. We then show the convergence of the resulting algorithm, max-DLT, to a stationary point of the DLT bound. 
Finally, we rely on extensive simulations of various network configurations, to establish the fast-converging nature of our proposed schemes, and thus their suitability for addressing the short coherence interval, as well as the increased system dimensions, arising when managing interference in lower bands of the millimeter wave spectrum. Moreover, our results also suggest that interference management still brings about significant performance gains, especially in dense deployments. 
\end{abstract}


\IEEEpeerreviewmaketitle

\section{Introduction}
Communication systems in the millimeter-wave (mmWave) band are one of the most promising candidate technologies for 5G systems, to address the ever-increasing demand for data-rates in cellular systems~\cite{Andrews_5G_14, Rappaport_millimeter_13}.
The systems we consider in this work operate at the lower bands of the mmWave frequency spectrum, sub-$28$ GHz systems, e.g., X-band ($8$-$12$) GHz, Ku-band ($12$-$18$) GHz, and $28$ GHz in the Ka band. 
The antenna spacing is not small enough to allow for hundreds of antennas, but rather a few tens (at most) at each transmitter/receiver. Thus, fully digital precoding is feasible, as the analog-to-digital converter power consumption is not a limiting factor.  
While the available bandwidth is narrower than higher mmWave frequency spectrum, sub-$28$ GHz systems offer several advantages over the latter: classical \emph{narrow-band} transmission/signal processing is feasible~\cite{Rappaprt_mmWprop}, channels follow Rayleigh/Rician fading in non line-of-sight environments~\cite{Rappaprt_mmWprop}, and pilot-based channel estimation is more suitable than beam alignment and channel sounding~\cite{Hur_mmWave_13,Alkhateeb_channel_2014}.
Investigations in the Ku band show that the narrow-band model is substantiated~\cite{Scalise_channelKu_08}. In that sense, they are \emph{transitional} architectures, between conventional LTE architectures (where interference management is critical), and future mmWave systems believed to be in the higher end of the spectrum (that are virtually interference-free). 

Interference management is less critical to mmWave communication, at $60$ GHz and beyond. Indeed outdoor links operating at $60$ GHz are shown to behave as \emph{pseudo-wired}, due to their highly directional nature\cite{Son_directmmW_12}. This results in channels whose sparsity (in terms of eigenmodes) is usually exploited for channel estimation\cite{Alkhateeb_channel_2014,Ghauch_SED_16}. 
In the system under consideration however, channels and beamforming are not highly directional (for a fixed array aperture), as compared to higher mmWave bands.
Moreover, the channels exhibit less \emph{sparsity} in non line-of-sight scenarios, than their counterparts that operate at higher frequencies: narrowband/wideband channel measurements over the $9.6$ GHz, $11.4$ GHz, and $28.8$ GHz bands, reveal that multipath components form a significant part of the received signal, in urban environments~\cite{Viotette_11GHz_88}. 

Thus, in such systems, when considering a multi-user multi-cell setup, interference is still a potentially limiting factor, and effective means of \emph{interference management} are still be needed. 
While several works have focused on coordination at the MAC layer (an exhaustive survey was done in~\cite{Niu_mmWsurvery_2015}), little-to-no work addresses the problem from a physical layer perspective. Performance evaluations of coordinated transmission at $28$ GHz, in a realistic propagation environment, reported gains in spectral efficiency - albeit moderate\cite{Biswas_compmmW_16}. 
Moreover, while earlier works such as~\cite{Rangan_millimeter_14} suggest that coordination and interference management bring about modest/little gains, for $28$ GHz systems, one has to also consider additional interference inherent to (ultra) dense deployments - a key feature of 5G systems~\cite{METISD62}.   
In addition, interference-limited scenarios arise due to intra-cell interference (as it is more stringent than inter-cell interference),
in the case of cell edge users, and/or when employing spatial multiplexing. In such cases, neglecting interference might be suboptimal. 
Shedding light on the above questions is central to this work.

Multi-user multi-cell coordination is often accomplished in an iterative distributed manner, where only local Channel State Information (CSI) is needed at each Base Station (BS) and Mobile Station (MS).
Such schemes employ Forward-Backward (F-B) iterations (also known as ping-pong iterations), to iteratively optimize the transmit and receive filters (Definition~\ref{def:F_B_training} in Sec.~\ref{sysmod}). 
Over the last decade, there has been a huge body of distributed coordination algorithms, for traditional multi-user multi-cell networks. 
Moreover, they can be categorized based on the metric that is optimized: interference leakage minimization~\cite{gomadam_distributed_2011,ghauch_interference_2011} and max-SINR~\cite{gomadam_distributed_2011}, minimum mean-squared error~\cite{schmidt_minimum_2009, peters_cooperative_2011}, weighted minimum mean-squared error~\cite{shi_wmmse_2011} and (weighted) sum-rate maximization~\cite{santamaria_maximum_2010, shi_wmmse_2011, Kaleva_SCPWSR_12}.

Despite the abundance of such schemes, they are \emph{ill-suited} for the problem at hand, as they require hundreds/thousands of iterations for convergence~\cite{Schmidt_comparison_13}. 
Furthermore, the number of required F-B iterations increases with the dimensions of the problem\cite{Schmidt_comparison_13}. They are thus only applicable to low-mobility scenarios, because the number of F-B iterations is limited by the coherence time of the channel. 
Note that the above limitations become stricter in the case of mmWave systems: more antennas at the transmitter and receiver are envisioned (and thus more F-B iterations until convergence), as well as lower coherence times compared to conventional sub-$6$ GHz systems (and thus a lower number of allowed F-B iterations) when the same mobility is assumed. 

Thus, applying such schemes to the sub-$28$ GHz systems under consideration, generates \emph{communication overhead} that offsets the resulting performance gains (as the communication overhead is dominated by the number of F-B iterations). 
Though this limitation is critical to coordination algorithms, just a handful of works have explored it, even in conventional sub-$6$ GHz systems. 
In line with recent work,~\cite{Komulainen_EffCSI_13, Nguyen_WSR_14, Brandt_FastConv_15, Ghauch_IWU_15}, investigating algorithms that operate in the 
low-overhead regime (where just a few F-B iterations are performed), is one of the main aims of this work. 
Moreover, our schemes are specifically designed to address the aforementioned limitations, by delivering superior performance under the low-overhead requirement, and increased system dimensions.    
With that in mind, while such schemes could equally well be applied to traditional cellular systems, their application has higher impact/relevance on the system at hand. 

We address the problem of sum-rate maximization in MIMO Interfering Multiple-Access Channels (MIMO IMAC), by formulating lower bounds on the problem. In a first part, we establish that maximizing the \emph{separability} between the signal subspace and the interference-plus-noise subspace (I+N), results in optimizing a bound on the sum-rate.
 In addition, we advocate the use of another separability metric, a lower bound on the sum-rate, that we refer to as a difference of log and trace (DLT) expression. We highlight the main advantages of using such an expression, namely that it yields optimization problems that decouple in both the transmit and receive filters. 
 We derive the solution to each of the subproblems - that we dub \emph{non-homogeneous waterfilling}, and underline its ability for stream control (by turning off streams that have low-SINR). We then propose a corresponding distributed algorithm, max-DLT, and establish its convergence to a stationary point of the DLT expression. 
Finally, we gear our numerical results to show the suitability of such schemes to the mmWave systems in question, by highlighting their fast-convergence and  superior performance, in the larger antenna regime. We also benchmark against several well-known schemes such as max-SINR~\cite{gomadam_distributed_2011}, (Weighted) MMSE~\cite{shi_wmmse_2011,peters_cooperative_2011}, and recent fast-converging approaches, such as CCP-WMMSE~\cite{Nguyen_WSR_14} and IWU~\cite{Ghauch_IWU_15}. 

Though the work addresses the problem at hand for a MIMO IMAC, it is equally applicable to the network-dual problem, the MIMO Interfering Broadcast Channel (IBC), and consequently to all the ensuing special cases, such as the MIMO IFC. This is further explored in the numerical results section. 

\emph{Notation:} we use bold upper-case letters to denote matrices, and bold lower-case denote vectors. For a given matrix $\pmb{A}$, we define $\tr(\pmb{A})$ as its trace, $\Vert \pmb{A} \Vert_F^2$ as its Frobenius norm, $|\pmb{A}|$ as its determinant,  $\pmb{A}^\dagger$ as its conjugate transpose, and $\bA^{-\dagger}$ as $(\bA^\dagger)^{-1}$.
 In addition, $\bA_{(i)}$ denotes its $i$th column, $\bA_{i:j} $ columns $i$ to $j$,  $\bA_{(i,j)}$ element $(i,j)$ in $\bA$, $\lambda_i[\pmb{A}]$ the $i^{th}$ eigenvalue of a Hermitian matrix $\pmb{A}$ (assuming the eigenvalues are sorted in decreasing order), and $v_{1:d}[\bA] $ denotes the $d$ dominant eigenvectors of $\bA$. 
$\mathbb{S}_{+}^{n,n}$ (resp. $\mathbb{S}_{++}^{n,n}$) is the set of complex $n \times n$ positive semi-definite (resp. positive definite) matrices.
Furthermore, $\pmb{A} \succ \pmb{0}$ (resp. $\pmb{A} \succeq \pmb{0}$) implies that $\pmb{A}$ is positive definite (resp. positive semi-definite), and $\pmb{A} \succ \pmb{B}$ (resp. $\pmb{A} \succeq \pmb{B} $) implies that $\pmb{A} - \pmb{B} \succ \ \pmb{0}$ (resp. $\pmb{A} - \pmb{B} \succeq \ \pmb{0}$).  
Finally, $\pmb{I}_n$ denotes the $n \times n$ identity matrix, $\lrb{n} = \lrb{1, \cdots, n } $, and $x^+ \triangleq \max \lrb{0, x}$.

\section{System Model and Problem Formulation} \label{sysmod}
\begin{figure}[!t]
	\center
	\includegraphics[ scale=.7]{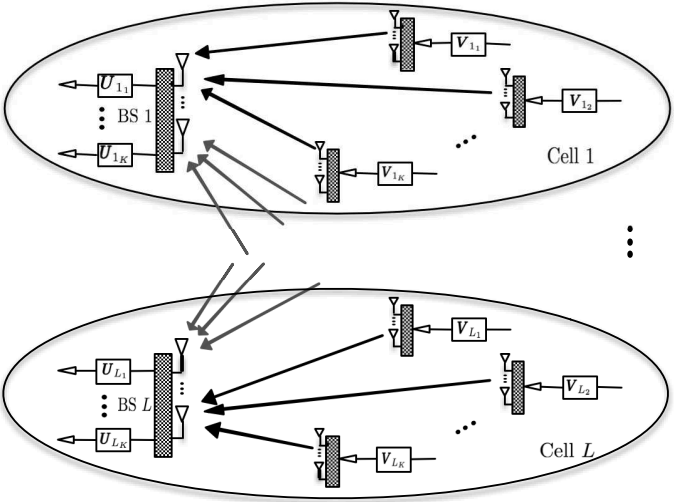}
	\caption{ $L$-cell MIMO Interfering Multiple-Access Channel} 
	\label{fig:setup}
\end{figure}

Consider a system with $L$ cells (each having one BS), where each cell is serving $K$ MSs (Fig.~\ref{fig:setup}). 
Each receiver (transmitter, resp.) is equipped with $N$ ($M$, resp.) antennas, and decodes $d$ data streams from each of its users ($d \leq \min(M,N)$). In the considered MIMO IMAC, transmitters (receivers, resp.) are MSs (BSs, resp.). Note that transmitters (receivers, resp.) become BSs (MSs, resp.), in the MIMO IBC scenario.
Let $\calL$ be the set of BSs, $\calK$ the set of users served by BS $l \in \calL $, 
and $l_j$ denote the index of user $j \in \calK$, at BS $l \in \calL$.
We denote by $\calI$ the total set of users, i.e., $\calI = \lrb{ l_j \ | \ l \in \calL , \ j \in \calK    } $.
The received signal at BS $l \in \calL $ is given by, 
\begin{align} \label{eq:sigmod}
\pmb{y}_l = \sum_{i \in \calL} \sum_{k \in \calK } \bH_{l, i_k} \bV_{i_k} \pmb{s}_{i_k} + \pmb{n}_l, \ l \in \calL ~,
\end{align}
To recover the signal of user $j \in \calK $, in cell $ l \in \calL$ (henceforth referred to as user $l_j \in \calI $),
$\by_l$ is processed with a linear filter, $\bU_{l_j} \in \IC^{N \times d} $, i.e., 
\begin{align}
\pmb{\tilde{s}}_{l_j} &=   \bU_{l_j}^\dagger \bH_{l, l_j} \bV_{l_j} \pmb{s}_{l_j} \nonumber \\
&~~~+ \sum_{\substack{ i \in \calL \\ i \neq l}} \sum_{  k \in \calK }  \bU_{l_j}^\dagger \bH_{l,i_k} \bV_{i_k} \pmb{s}_{i_k} +  \bU_{l_j}^\dagger \pmb{n}_l, \ \forall \ l_j \in \calI
\end{align}
where the first term represents the desired signal, the second both the intra and inter-cell interference. In the above, $\pmb{s}_{i_k}$ represents the $d$-dimensional vector of \emph{independently encoded symbols} for user $i_k \in \calI$, with covariance matrix $\mathbb{E}[\pmb{s}_{i_k} \pmb{s}_{i_k}^\dagger ] =  \pmb{I}_d $. In addition, $\bV_{i_k}$ denotes the $M \times d$ transmit filter of user $i_k \in \calI$, and $\bH_{l, i_k} $ the $N \times M$ MIMO channel from user $i_k \in \calI$, to BS $l$ (assumed to be block-fading with i.i.d. entries). $\pmb{n}_l$ represents the $N$-dimensional AWGN noise at BS $l \in \calL$, such that $\mathbb{E}[\pmb{n}_l \pmb{n}_l^\dagger ] = \sigma_l^2 \pmb{I}_N $. Note that our model (and the results presented thereafter) can easily be extended to cases where $M, N$ and $d$ are different across users and BSs.

If we assume that joint encoding/decoding of each user's streams is performed at the users and BSs, and treating interference as noise, the achievable rate of user $l_j \in \calI $ is given by,
\begin{align}
r_{l_j} = \log_2 | \pmb{I}_d + (\bU_{l_j}^\dagger \bR_{l_j} \bU_{l_j}) (\bU_{l_j}^\dagger \bQ_{l_j} \bU_{l_j})^{-1} |, \ l_j \in \calI  \label{eq:user_rate}
\end{align}
where $\bR_{l_j}$ and $\bQ_{l_j}$ are the desired signal and interference-plus-noise (I+N) covariance matrices for user $j$, at BS $l$, respectively, and are given by,
\begin{align*}
\bR_{l_j} &=  \bH_{l, l_j} \bV_{l_j} \bV_{l_j}^\dagger \bH_{l , l_j}^\dagger , l_j \in \calI  \\
\bQ_{l_j} &= \sum_{i=1}^L \sum_{k=1}^K \bH_{l, i_k}\bV_{i_k} \bV_{i_k}^\dagger \bH_{l,i_k}^\dagger + \sigma_l^2 \pmb{I}_N - \bR_{l_j},  l_j \in \calI .
\end{align*}  
 Moreover, we define
\begin{align*}
 \bar{\bR}_{i_k} &=  \bH_{i , i_k}^\dagger \bU_{i_k} \bU_{i_k}^\dagger \bH_{i, i_k} , \  i_k \in \calI  \\
 \bar{\bQ}_{i_k} &= \sum_{l=1}^L \sum_{j=1}^K \bH_{l , i_k}^\dagger \bU_{l_j}  \bU_{l_j}^\dagger \bH_{l, i_k} + \bar{\sigma}_{i_k}^2 \pmb{I}_M - \bar{\bR}_{i_k}, i_k \in \calI
\end{align*}
as the signal and I+N covariance matrices of user $i_k$, in the reverse network (where $ \bar{\sigma}_{i_k}^2 $ is the noise variance at user $i_k$). Finally, we henceforth denote $\bL_{l_j} \bL_{l_j}^\dagger$ as the Cholesky Decomposition of $\bQ_{l_j}$, and $\bK_{i_k} \bK_{i_k}^\dagger$ as that of $\bar{\bQ}_{i_k}$. We formulate the (unweighted) sum-rate maximization problem as follows,
\begin{align}
\max R_{\Sigma} (\lbrace \bU_{l_j} \rbrace, \lbrace \bV_{l_j} \rbrace) \triangleq \sum_{l \in \calL } \sum_{j \in \calK } r_{l_j} \label{jsrm} ~.
\end{align}
In the next section we generalize the well-known max-SINR algorithm from a stream-by-stream optimization algorithm, to an algorithm that optimizes the whole transmit/receive filter. For that purpose, we show that this generalized form can be formulated using \emph{separability} metrics, namely, the Generalized Multi-dimensional Rayleigh Quotient (GMRQ), defined next. We next highlight the central assumptions/definitions of this work. 

\subsection{Preliminaries }	
\begin{figure*}
	\center
	\includegraphics[ scale=.75]{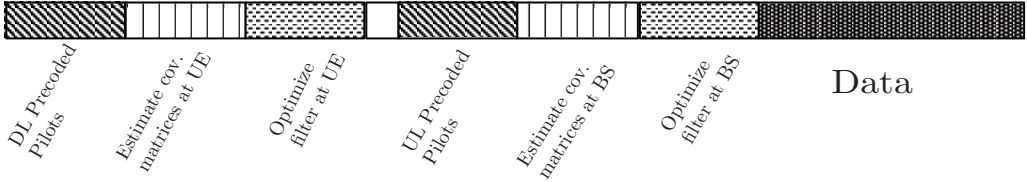}
	\caption{ Basic structure of Forward-Backward Iteration } 
	\label{fig:frame}
\end{figure*}

The schemes we consider in the present work fall under the category of Forward-Backward training, recapped in the definition below. 
\begin{definition}[F-B Training] \label{def:F_B_training}  \rm
Schemes employing Forward-Backward (F-B) iterations (also known as ping-pong iterations, or bi-directional training), consist of optimizing the receive filters (at the BSs) in the forward training phase, then the transmit filters (at the MSs) in the reverse training phase. They exploit channel reciprocity in Time-Division Duplex (TDD) systems, and result in fully distributed algorithms. The basic iteration structure is shown in Fig.~\ref{fig:frame}.
\end{definition}

\begin{definition}[Separability] \label{def:separ}  \rm
Given two sets of points with covariance matrices $\bR$ and $\bQ$, separability is a measure of the distance between the sets, after projecting on a subspace $\bU$. 
Separability metrics - the building blocks of areas such as linear discriminant analysis~\cite[Chap. 4.1]{Bishop_PatternRec_06}, include 
the \emph{Generalized Multi-dimensional Rayleigh Quotient}: 
	$$ ~~~~\text{GMRQ} = \frac{|\bU^\dagger \bR \bU|}{ |\bU^\dagger \bQ \bU|} $$ 
In the context of this work, $\bR$ and $\bQ$ represent the signal and I+N covariance matrices, respectively, and $\bU$ the linear filter at the receiver.
\end{definition}

\begin{assumption}[Local CSI] \rm
We assume that each MS/BS has local CSI, i.e., each MS (resp. BS) knows the channels to its desired and interfering BSs (resp. users). 
Although we underline that methods in~\cite{Brandt_DistCSI_15} are applicable for acquiring such quantities (discussion in Sect.~\ref{sec:dist_csi_acqusition}), investigating the CSI acquisition mechanism is not part of this work. Moreover, local CSI at each MS/BS is assumed to be perfectly known. 
\end{assumption}
\begin{assumption}[Distributed Operation] \rm
All schemes are required to use local CSI only, using the framework of F-B training. 
\end{assumption}
\begin{assumption}[Low-Overhead Regime] \rm
We restrict our proposed schemes to operate in the low-overhead regime, where only a small number of F-B iterations is used
(in line with recent work such as~\cite{Komulainen_EffCSI_13, Brandt_FastConv_15, Ghauch_IWU_15,Nguyen_WSR_14}).  
\end{assumption}

\section{Separability and Sum-rate maximization} \label{sec:aims}
\subsection{Problem Formulation}
In this part, we shed light on the intimate relation between sum-rate maximization and maximization of the GMRQ separability metric.
We make use of the fact that $\log|\bX|$ is monotonically increasing on the positive-definite cone, i.e., 
  $\log|\bX_2| \geq \log| \bX_1 | , \ \text{for } \ \bX_2 \succeq \bX_1 \succ \pmb{0} $. 
Applying the above property, we lower bound $r_{l_j}$ in~\eqref{eq:user_rate} as, 
\begin{align}
r_{l_j} &> \log_2|( \bU_{l_j}^\dagger \bR_{l_j} \bU_{l_j})( \bU_{l_j}^\dagger \bQ_{l_j} \bU_{l_j})^{-1}|  \nonumber \\
&= \log_2 \frac{|  \bU_{l_j}^\dagger \bR_{l_j} \bU_{l_j} | }{|  \bU_{l_j}^\dagger \bQ_{l_j} \bU_{l_j} |} \triangleq \tilde{r}_{l_j} , \ \forall l_j \in \calI  \label{eq:ratelb}
\end{align} 
Note that $\tilde{r}_{l_j}$ is a high-SNR approximation of the actual user rate $r_{l_j}$, where the approximation error is $\calO(\tr[(\bU_{l_j}^\dagger \bQ_{l_j} \bU_{l_j})(\bU_{l_j}^\dagger \bR_{l_j} \bU_{l_j})^{-1}  ])$ (refer to Appendix~\ref{app:dlt_bound}). Moreover, bounds such as~\eqref{eq:ratelb} are already prevalent in the MIMO literature (e.g.,~\cite[Proposition 2]{Oyman_MIMOCapapprox_03}). 
Thus, the sum-rate $R_{\Sigma}$ can be bounded below, as follows, 
\begin{align*}
R_{\Sigma} &> \sum_{l_j \in \calI } \tilde{r}_{l_j} = \log_2 ( \prod_{l_j}  q_{l_j} ) , \ \textrm{where } q_{l_j} \triangleq \frac{|  \bU_{l_j}^\dagger \bR_{l_j} \bU_{l_j} | }{|  \bU_{l_j}^\dagger \bQ_{l_j} \bU_{l_j} |} 
\end{align*} 
Since $\log(-)$ is monotonic, the sum-rate maximization problem,~\eqref{jsrm}, is lower bounded by, 
\begin{align}
(SRM)
\begin{cases} 
         \underset{ \lbrace \bU_{l_j} , \bV_{l_j} \rbrace}{\max}  \prod_{l_j \in \calI } \ q_{l_j}   \\
         \st  \ \Vert \bU_{l_j} \Vert_F^2 = P_r \ ,   \Vert \bV_{l_j} \Vert_F^2 = P_t \ ,   \forall \ l_j \in \calI
\end{cases}
\end{align}

\begin{remark}[Power Constraint] \rm
In distributed optimization schemes employing Forward-Backward (F-B) training, receivers are active in one of the phases (i.e., by sending data / pilots). Thus, generally, in the forward training phase (Definition~\ref{def:F_B_training}), one needs a maximum transmit power constraint for the receiver filter (at each BS). This is in addition to the maximum transmit power constraint of the transmitter (at each MS), widely used in cellular systems. In addition, in scenarios involving a multi-cellular downlink communication, each BS employs a sum-power constraint for its users, e.g.,~\cite{shi_wmmse_2011}. However, the same does not hold in the considered setup (multi-cellular uplink), since it would lead to a sum-power constraint, across all MSs: clearly this is not applicable in practice.
We thus adopt an individual per-user power constraint. We also assume equal power allocation among the users in a cell, to avoid the need for power allocation (outside the scope of this work).
From a mathematical perspective however, the cost function in $(SRM)$ renders the presence of a receive power constraint, irrelevant. 
\end{remark}

Referring to $(SRM)$, $q_{l_j}$ is nothing but the GMRQ separability metric in Definition~\ref{def:separ}. 
Consequently, given the signal and I+N covariance matrices, $\bR_{l_j}$ and $\bQ_{l_j}$, each receiver chooses its filter such as to maximize the separation between signal and I+N subspaces.

\subsection{Maximization of GMRQ}
The main limitation of solving problems such $(SRM)$ is the fact it is not jointly convex in all the optimization variables. Though Block Coordinate Decent (BCD) stands out as a strong candidate, one major obstacle persists: while the problem decouples in the receive filters (as shown in $(SRM)$), attempting to write a similar expression by factoring out the transmit filters, leads to a coupled problem. Therefore, we propose an alternative (purely heuristic) method: the receive filters are updated as the solution to maximizing the sum-rate (assuming fixed transmit filters), while the transmit filters are chosen as the solution of the reverse network sum rate maximization (this same structure is implicitly exploited in max-SINR~\cite{gomadam_distributed_2011}), i.e., 
\begin{align*}
(SRM_F) 
\begin{cases}
 	\underset{ \lbrace \bU_{l_j} \rbrace }{\max} \ \prod_{l_j \in \calI } \ \ q_{l_j} (\bU_{l_j}) =  \frac{|  \bU_{l_j}^\dagger \bR_{l_j} \bU_{l_j} | }{|  \bU_{l_j}^\dagger \bQ_{l_j} \bU_{l_j} |}   \\
    \st \ \Vert \bU_{l_j} \Vert_F^2 = P_r  \ ,  \ \forall \ l_j  \in \calI \ ,
\end{cases}
\textup{and}, 
\end{align*}

\begin{align*}
(SRM_B) 
\begin{cases}
       \underset{ \lbrace \bV_{i_k} \rbrace }{\max}  \prod_{i_k \in \calI } \ \ p_{i_k} (\bV_{i_k}) =  \frac{| \bV_{i_k}^\dagger \bar{\bR}_{i_k} \bV_{i_k} | }{| \bV_{i_k}^\dagger \bar{\bQ}_{i_k} \bV_{i_k} |} \\
       \st \ \Vert \bV_{i_k} \Vert_F^2 = P_t  \ ,  \ \forall \ i_k \in \calI .
\end{cases}
\end{align*}
In other words, assuming transmit filters as fixed, the receive filters are updated such as to maximize the separability metric in the forward phase. Similarly, the transmit filters are chosen to maximize the separability in the backward training phase. Moreover, as seen from the above problems, the objective in each subproblem is invariant to scaling of the optimal solution. Thus, they can be solved as unconstrained problems, without loss of optimality.   

We first require a solution to the GMRQ maximization. The solution was earlier proposed in~\cite{prieto_generalized_2003}, and is not a contribution of this work. It is restated below for completeness. 
\begin{lemma} \label{lem_grq}
Consider the following maximization of the $r$-dimensional GMRQ, 
\begin{align}
&\bX^\star \triangleq  \underset{ \bX \in \mathbb{C}^{n \times r} }{\argmax} \ q(\bX) = \frac{| \bX^\dagger \bR \bX  | }{| \bX^\dagger \bQ \bX |} \ ,  \label{gmrq}
\end{align}
where $ \bQ \in \IS_{++}^{n \times n} \ , \ \ \bR \in \IS_{+}^{n \times n}$ and $r < n$. The set of optimal solutions to this non-convex problem are given by, 
\begin{align}
\bX^\star =  \pmb{L}^{-\dagger} \bPsi \hat{\bV} \ ,  \label{gmrqsol}
\end{align}
where $\pmb{LL}^\dagger = \bQ  \ (\bL \in \IC^{n \times n}) $, 
$\bPsi = v_{1:r}[\pmb{L}^{-1} \bR \pmb{L}^{-\dagger} ]  \  (\bPsi \in \IC^{n \times r})$, and $\hat{\bV} \in \IC^{r \times r} $ is arbitrary and non-singular. 
\end{lemma}
\begin{IEEEproof} 
It was shown in~\cite{prieto_generalized_2003} that a solution to (\ref{gmrq}) is given by, $ \bX^\star  =  \bL^{-\dagger} \bPsi $. 
We note that it can verified that this optimal solution is invariant to multiplication by a non-singular matrix $\hat{\bV}$,  i.e.,  $q( \bX^\star \hat{\bV} ) = q(\bX^\star ) $. Thus, a generic form of the solution is, $\bX^\star  =\bL^{-\dagger} \bPsi \hat{\bV}$
\end{IEEEproof}
Note that the above solution is a generalized formulation of the well-known generalized eigenvalues solution. This  equivalence was also established in in~\cite{prieto_generalized_2003}, and is restated below for convenience. 

\begin{corollary}\label{cor_grq}
Consider a special case of~\eqref{gmrqsol} where $ \hat{\pmb{V}} = \bI_r $. Then, this corresponds to the generalized eigenvalues solution, 
\begin{align}
\bX^\star =  \pmb{L}^{-\dagger} \bPsi \Leftrightarrow \bR \bX^\star = \bQ \bX^\star \bLam_r 
\end{align}
 where  $\bLam_r \in \mathbb{R}^{r \times r} $  be the (diagonal) matrix of eigenvalues for $\pmb{L}^{-1} \bR \pmb{L}^{-\dagger}$.
\end{corollary} 
\begin{IEEEproof} Refer to~\cite{prieto_generalized_2003}.    
\end{IEEEproof}

With this in mind, we can write the optimal transmit and receive filter updates, as follows, 
\begin{align}
\bU_{l_j}^\star &=  \bL_{l_j}^{-\dagger} \bPsi_{l_j}, \ \ \bPsi_{l_j} \triangleq  v_{1:d} [ \bL_{l_j}^{-1} \bR_{l_j} \bL_{l_j}^{-\dagger} ]  \ , \ \forall \ l_j \ , \nonumber \\
\bV_{i_k}^\star &=  \bK_{i_k}^{-\dagger} \bTh_{i_k}, \ \ \bTh_{i_k} \triangleq v_{1:d} [ \bK_{i_k}^{-1} \bar{\bR}_{i_k} \bK_{i_k}^{-\dagger}]  \ ,  \ \forall i_k \ ,   \label{upt}
\end{align}
where we used the fact we can set $\hat{\bV} = \pmb{I}_d$ in the solution of (\ref{gmrqsol}). 
We note that the optimal filter updates for the transmitter are more heuristic than the receiver ones: 
While the receive filter updates directly maximize a lower bound on the sum-rate - as seen in $(SRM)$, no such claim can be made about the transmit filter updates.   
The details of our algorithm, Alternating Iterative Maximal Separation (AIMS), are shown in Algorithm~\ref{alg1} (where $T$ denotes the number of F-B iterations).

\begin{algorithm} 
\caption{Alternating Iterative Maximal Separation (AIMS)} \label{alg1}
\begin{algorithmic}
\For{$t=1,2,..., T$}
\State // \emph{forward network optimization: receive filter update}
\State \hspace{.2cm} Estimate   $\bR_{l_j}, \bQ_{l_j}$, and compute $\bL_{l_j}$, $\forall l_j$
\State \hspace{.2cm}  $\bU_{l_j} \leftarrow  \bL_{l_j}^{-\dagger}  v_{1:d} [ \bL_{l_j}^{-1} \bR_{l_j} \bL_{l_j}^{-\dagger} ], \ \ \forall l_j$
\State \hspace{.2cm} $\bU_{l_j}  \leftarrow  \sqrt{P_r} \ \bU_{l_j}  /  \Vert \bU_{l_j} \Vert_F $
\State // \emph{reverse network optimization: transmit filter update}
\State \hspace{.2cm} Estimate   $\bar{\bR}_{i_k}, \bar{\bQ}_{i_k}$, and compute $\bK_{i_k}$ ,  $\forall i_k$
\State \hspace{.2cm} $\bV_{i_k} \leftarrow  \bK_{i_k}^{-\dagger}  v_{1:d} [ \bK_{i_k}^{-1} \bar{\bR}_{i_k} \bK_{i_k}^{-\dagger} ] , \ \ \forall i_k$
\State \hspace{.2cm} $\bV_{i_k}  \leftarrow  \sqrt{P_t} \ \bV_{i_k}  /  \Vert \bV_{i_k} \Vert_F $
\EndFor 
\end{algorithmic}
\end{algorithm} 

Using the above solution, we next establish the result that employing unitary filters is not optimal, from the perspective of separability. 
We stress that the latter is not central to the main story of this work, but rather an interesting result from the separability perspective, that is obtained `for free'. We thus restrict our presentation to sketching a proof, in Appendix~\ref{A3a}.   
\begin{proposition} \label{prop:nonunitary}
Consider the optimal receive filter given in (\ref{upt}), i.e., $\bU_{l_j}^\star =  \bL_{l_j}^{-\dagger} \bPsi_{l_j} $, where $\bPsi_{l_j} = v_{1:d} [ \bL_{l_j}^{-1}\bR_{l_j} \bL_{l_j}^{-\dagger} ]  $. Then, assuming MIMO channel coefficients are i.i.d. (as defined in Sec. II), $\bU_{l_j}^\star $ is not orthonormal, almost surely. 
\end{proposition} 
A few comments are in order at this stage, regarding the difference between AIMS and max-SINR. Referring to $(SRM_F)$ and $(SRM_B)$, it is clear that our proposed algorithm reduces to max-SINR, in case of single-stream transmission, i.e., setting $d=1$. Moreover, an inherent property of the max-SINR solution is that it yields equal power allocation across all the streams (since the individual columns of each transmit/receive filter are normalized to unity). 
However, as evident from~\eqref{upt}, our proposed solution does not normalize the individual columns of the receive filter, but rather the whole filter norm (as seen in Algorithm~\ref{alg1}). 
This allows for different power allocation, across columns of the same filter.  
That being said, the proposed solution is expected to yield better sum-rate performance (w.r.t. max-SINR), especially in the interference-limited regime. This is due to the intuitive fact that much can be gained from allocating low power to streams that suffer from severe interference, and higher power to streams with lesser interference (this will be validated in the numerical results section). We next introduce a rank adaptation mechanism that further enhances the interference suppression capabilities of the algorithm. 

\subsection{AIMS with Rank Adaptation} \label{sec:RankAdapt}
We introduce an additional (heuristic) mechanism to robustify AIMS against severely interference-limited scenarios, by introducing a mechanism of Rank Adaptation (RA): in addition to the transmit / receive filter optimization (Lemma~\ref{lem_grq}), the latter allows the filter rank to be optimized as well. 
Mathematically speaking, RA addresses the following problem, 
\begin{align}
& r^\star \triangleq  \underset{ r }{\argmax} \left[ \bX^\star \triangleq  \underset{ \bX \in \mathbb{C}^{n \times r} }{\argmax} \frac{| \bX^\dagger \bR \bX  | }{| \bX^\dagger \bQ \bX |} \right]   , \label{opt:gmrq_ra}
\end{align}
Using the same argument as Lemma~\ref{lem_grq}, one can verify that $\bX^\star$ and $r^\star$ are as follows, 
\begin{align}
&\bX^\star =  [\pmb{L}^{-\dagger} \bPsi]_{1:r^\star } \ ,  \ \textrm{where} \ \bPsi = v_{1:n}[\pmb{L}^{-1} \bR \pmb{L}^{-\dagger} ] \nonumber \\  
& r^\star = \underset{ r }{\argmax} \ | \bLam_r | = \left| \lrb{  i \ |  \ \lambda_i[ \pmb{L}^{-1} \bR \pmb{L}^{-\dagger}  \geq 1  } \right| 
\end{align}
where  $\bLam_r \in \mathbb{R}^{r \times r} $  is the (diagonal) matrix consisting of the $r$-largest eigenvalues of $\pmb{L}^{-1} \bR \pmb{L}^{-\dagger}$. Simply put, $r^\star$ is the number of eigenvalues greater than one. 

When RA is incorporated into AIMS, this mechanism will boost the performance of the algorithm (namely in interference-limited settings). 
However, one still needs to ensure that the filter ranks for each transmit-receive pair are the same, i.e., $ \textrm{rank} (\bU_{l_j}) = \textrm{rank} (\bV_{l_j}) \ \forall \ l_j $. One quick (heuristic) solution is as follows. For each transmit-receive filter pair, compute the optimal filter rank for both the transmit and receive filter, and use the minimum.\footnote{
Alternately, one can apply RA to the receive filters only, in the last iteration of the algorithm, since the transmit filter updates are more heuristic than the receive filter updates.}  
Needless to say, ensuring this condition requires additional signaling overhead. We thus envision RA, as potential `add-on' for AIMS, when one can afford the resulting overhead increase. The added performance boost from RA in further discussed in the numerical results section.

\section{Maximizing a DLT bound}
In this section we propose another approach to tackle the sum-rate optimization problem. The central idea behind this approach is to use a lower bound on the sum-rate, that results in separable subproblems. 
\subsection{Problem Formulation}
We focus the derivations to the interference-limited case,  
where the following holds,
\begin{align} 
&\lambda_i[\bU_{l_j}^\dagger \bQ_{l_j} \bU_{l_j}] \rightarrow \infty , \  \forall i \in \lrb{d} \nonumber \\
&~~\Leftrightarrow 
\begin{cases}
A1) \ \lambda_i[(\bU_{l_j}^\dagger \bQ_{l_j} \bU_{l_j})^{-1}] \rightarrow 0 ,  \ \forall i \in \lrb{d}   \\
A2) \ \bI_d  \succeq (\bU_{l_j}^\dagger \bQ_{l_j} \bU_{l_j})^{-1} \label{eq:intf_limited} \\
\end{cases}
\end{align}
\begin{proposition}\label{prop:dlt_bound}
In the interference-limited regime, the user-rate $r_{l_j}$ in~\eqref{eq:user_rate} is lower bounded by, 
\begin{align}
r_{l_j} &\geq \log_2 | \bI_d + \bU_{l_j}^\dagger \bR_{l_j} \bU_{l_j} |  - \log_2 |  \bU_{l_j}^\dagger \bQ_{l_j} \bU_{l_j} | ,  \label{eq:DLL_lb} \\
&\geq \log_2 | \pmb{I}_d +  \bU_{l_j}^\dagger \bR_{l_j} \bU_{l_j} |  - \tr(  \bU_{l_j}^\dagger \bQ_{l_j} \bU_{l_j} ) \triangleq r_{l_j}^{(LB)},  \label{eq:DLT_lb}
\end{align}
where $r_{l_j}^{(LB)}$ is such that, 
\begin{align}
\Delta_{l_j}  &\triangleq r_{l_j} - r_{l_j}^{(LB)} \nonumber \\
&=  \tr(  \bU_{l_j}^\dagger \bQ_{l_j} \bU_{l_j} ) - \log_2|\bU_{l_j}^\dagger \bQ_{l_j} \bU_{l_j} |  \nonumber \\
&~~+ \calO(\tr[(\bU_{l_j}^\dagger \bQ_{l_j} \bU_{l_j})(\bU_{l_j}^\dagger \bR_{l_j} \bU_{l_j})^{-1}  ]) ,  \ \forall l_j \in \calI \
\end{align}
\end{proposition}
\begin{IEEEproof} Refer to Appendix~\ref{app:dlt_bound}. 
\end{IEEEproof}
We refer to expressions such as $r_{l_j}^{(LB)}$, as a Difference of Log-Trace (DLT) expressions. They shall be used as basis for the optimization algorithm. 
With that in mind, the sum-rate $R_{\Sigma}$, can be lower bounded by $R_{\Sigma}^{(LB)}$, 
\begin{align}
R_{\Sigma}^{(LB)} 
&= \sum_{l_j \in \calI} \log_2 | \pmb{I}_d +  \bU_{l_j}^\dagger \bR_{l_j} \bU_{l_j} |  - \tr(  \bU_{l_j}^\dagger \bQ_{l_j} \bU_{l_j} )  \label{eq:rate}
\end{align}
\begin{align}
&= \sum_{i_k \in \calI} \log_2 | \pmb{I}_d + \bV_{i_k}^\dagger \bar{\bR}_{i_k} \bV_{i_k} |  - \tr( \bV_{i_k}^\dagger \bar{\bQ}_{i_k} \bV_{i_k} ) \label{eq:raterev}
\end{align}
where the last equality is due to $\log| \bI + \bA\bB | = \log| \bI + \bB \bA | $, and the linearity of  $\tr(.)$. 
Then, the sum-rate optimization problem in~\eqref{jsrm} can be bounded below by solving the following, 
\begin{align} 
\begin{cases}  \label{opt:smrlb}
         \underset{ \lrb{\bV_{l_j},\bU_{l_j}} }{\max}  R_{\Sigma}^{(LB)}  \\
         \st \ \Vert \bU_{l_j} \Vert_F^2 = P_r , \  \Vert \bV_{l_j} \Vert_F^2 = P_t , \ \forall l_j \in \calI
\end{cases} 
\end{align}
Note that the above problem is not jointly convex in all the optimization variables, mainly due to the coupling between the transmit and receive filters. 
Moreover, the reason for having transmit/receive power constraints with equality, will become clear in Sec.~\ref{sec:discussion}.

\subsection{Proposed Algorithm}
The formulation in~\eqref{opt:smrlb} is ideal for a Block Coordinate Descent (BCD) approach. 
We use the superscript $^{(n)}$ to denote the iteration number: at the $n$th iteration, the transmit filters, $\lrb{\bV_{l_j}^{(n)}}$, are fixed, and the update for the receive filters, $\lrb{\bU_{l_j}^{(n+1)}}$, is the one that maximizes the objective. The same is done for the transmit filter update. 
In each of the two stages, BCD decomposes the original coupled problem~\eqref{opt:smrlb}, into a set of parallel subproblems, that can solved in distributed fashion. This is formalized in~\eqref{opt:bcd}, and each of the resulting subproblems are detailed below. 
\begin{figure*}
\begin{align} \label{opt:bcd}
 \underbrace{ \lrb{ \bV_{l_j}^{(n+1)} } \triangleq \underset{ \lrb{ \bV_{l_j} } }{\argmax} \ R_{\Sigma}^{(LB)} \left( \underbrace{ \lrb{\bU_{l_j}^{(n+1)}} \triangleq \underset{ \lrb{ \bU_{l_j} } }{\argmax} \ R_{\Sigma}^{(LB)} ( \lrb{\bU_{l_j}} , \lrb{\bV_{l_j}^{(n)}} ) \ }_{J1}, \  \lrb{\bV_{l_j}}  \right) }_{J2}   , \ n = 1,2,...
\end{align} 
\hrule \normalsize
\end{figure*}
When the transmit filters are fixed, the problem decouples in the receive filters $\lrb{\bU_{l_j}}$ (as seen from~\eqref{opt:smrlb}), and the resulting subproblems are given by,
\begin{align} 
(J1) \ 
\begin{cases}  \label{opt:rlbrx}
         \underset{ \bU_{l_j} }{\min} \ \tr(  \bU_{l_j}^\dagger \bQ_{l_j} \bU_{l_j} ) -  \log_2 | \pmb{I}_d +  \bU_{l_j}^\dagger \bR_{l_j} \bU_{l_j} |   \\
         \st \ \Vert \bU_{l_j} \Vert_F^2 = P_r 
\end{cases} 
\end{align}
By recalling that~\eqref{opt:smrlb} can rewritten as~\eqref{eq:raterev}, we see that the above objective decouples in the transmit filters, i.e., 
\begin{align} 
(J2) \
\begin{cases}  \label{opt:rlbtx}
         \underset{ \bV_{i_k} }{\min} \ \tr( \bV_{i_k}^\dagger \bar{\bQ}_{i_k} \bV_{i_k} ) -  \log_2 | \pmb{I}_d + \bV_{i_k}^\dagger \bar{\bR}_{i_k} \bV_{i_k} |   \\
         \st \ \Vert \bV_{i_k} \Vert_F^2 = P_t
\end{cases} 
\end{align}
Thus, choosing  DLT expressions is rather advantageous, since they lead to subproblems that decouple in both $\lrb{\bU_{l_j}}$ and $\lrb{\bV_{l_j}}$. 
Note that the equality constraints in $(J1)$ and $(J2)$, do not affect the convexity of the problems, as they are already non-convex. Indeed, expressions such as  $-\log_2 | \pmb{I}_d +  \bU_{l_j}^\dagger \bR_{l_j} \bU_{l_j} | $ are \emph{not convex} in $\bU_{l_j}$.
\footnote{To see this, consider the (degenerate) scalar case. It is easily verified that $-\log_2(1 + r u^2 ) , \ r > 0 $ is concave for $u \ll 1 $, and convex for $u \gg 1 $.}
However, this does not make BCD less applicable, as long as $(J1)$ and $(J2)$ are solved \emph{globally}. 
The solution to each of the subproblems is given by the following result. 
\begin{lemma}  \label{lem_dlt} 
Non-homogeneous Waterfilling. \\*
Consider the following problem,
\begin{align} 
(P) \ \begin{cases}  \label{opt:logtr}
         \underset{\bX \in \mathbb{C}^{n \times r} }{\min} \ f(\bX) \triangleq  \tr( \bX^\dagger \bQ \bX ) - \log_2 | \pmb{I}_d + \bX^\dagger \bR \bX |   \\
         \st \ \Vert \bX \Vert_F^2 = \zeta .
\end{cases}
\end{align}
where $ \bQ \succ \pmb{0} $ and $\bR \succeq \pmb{0}, \ r < n$. Let $\bQ \triangleq \bL\bL^\dagger$ be the Cholesky factorization of $\bQ$, and $\bM \triangleq \bL^{-1} \bR \bL^{-\dagger}, \ \bM \succeq \pmb{0} $, and define the following,
 $\lrb{ \alpha_i \triangleq \lambda_i[\bM] }_{i=1}^r $ , $ \ \bPsi \triangleq v_{1:r}[\bM]$, $ \lrb{ \beta_i \triangleq \bPsi_{(i)}^\dagger (\bL^\dagger \bL)^{-1}  \bPsi_{(i)} }_{i=1}^r $. Then the optimal solution for $(P)$ is given by, 
\begin{align} \label{eq:nhwf}
\bX^\star = \bL^{-\dagger} \bPsi \bSig^\star ,
\end{align}
where $\bSig^\star$ (diagonal) is the optimal power allocation, 
\begin{align} \label{opt:pow_alloc}
(P5) \ 
\begin{cases}
\underset{ \lrb{x_i} }{\min} \sum_{i=1}^r \left( x_i - \log_2( 1 + \alpha_i x_i ) \right) ~ \\
\st \ \sum_{i=1}^r \beta_i x_i = \zeta, \ x_i \geq 0, \forall i  
\end{cases}
\end{align}
\end{lemma}
\begin{IEEEproof} Refer to Appendix~\ref{A1} \end{IEEEproof}
We underline that a similar problem was obtained in~\cite{Kim_adhocCR_11}, but in the context of covariance optimization. Hence, this result is not applicable to $(P)$. Moreover, $(P5)$ has a closed-form solution, that can be obtained using standard Lagrangian techniques. 
\begin{lemma} \label{lem:optpow_alloc}
The solution to the optimal power allocation in $(P5)$ is given by, 
\begin{align} \label{eq:optpow}
\bSig_{(i,i)}^\star = \sqrt{\Big( 1/(1+\mu^\star \beta_i) - 1/\alpha_i  \Big)^+}, \forall i ,
\end{align}
where $\mu^\star$ is the unique root to,
$$g(\mu) \triangleq \sum_{i=1}^r \beta_i \Big( 1/(1+\mu \beta_i) - 1/\alpha_i  \Big)^+  - \zeta ,$$
on the interval $] -1/(\max_i \beta_i ) , \ \infty [ $, and $g(\mu)$ is monotonically decreasing on that interval. 
\end{lemma}
 \begin{IEEEproof} Refer to Appendix~\ref{A1} \end{IEEEproof}

With this in mind, we can write the optimal transmit and receive filter updates, as follows, 
\begin{align}
\bU_{l_j}^\star &=  \bL_{l_j}^{-\dagger} \bPsi_{l_j} \  \bSig_{l_j}^\star , \ \   \bPsi_{l_j} \triangleq v_{1:d} [ \bL_{l_j}^{-1} \bR_{l_j} \bL_{l_j}^{-\dagger} ], \forall \ l_j \ , \nonumber \\  
\bV_{i_k}^\star &=  \bK_{i_k}^{-\dagger} \bTh_{i_k} \  \bLam_{i_k}^\star , \ \   \bTh_{i_k} \triangleq v_{1:d} [ \bK_{i_k}^{-1} \bar{\bR}_{i_k} \bK_{i_k}^{-\dagger} ], \forall \ i_k \ , \label{eq:upt2}
\end{align}
where $ \bSig_{l_j}^\star$ and $ \bLam_{i_k}^\star$ are the optimal power allocation, given in Lemma~\ref{lem:optpow_alloc}. The resulting algorithm, max-DLT, is detailed in Algorithm~\ref{alg2} (where $T$ is the number of F-B iterations). 
Moreover, due to the monotone nature of $g(\mu)$, $\mu^\star$ can be found using simple 1D search methods.  

\begin{algorithm} 
\caption{Maximal DLT (max-DLT)} \label{alg2}
\begin{algorithmic}
\For{$t=1,2,..., T$}
\State // \emph{forward network optimization: receive filter update}
\State \hspace{.2cm} Estimate   $\bR_{l_j}, \bQ_{l_j}$, and compute $\bL_{l_j}$,    $\forall l_j$
\State \hspace{.2cm} $\bU_{l_j} \leftarrow  \bL_{l_j}^{-\dagger}  v_{1:d} [ \bL_{l_j}^{-1} \bR_{l_j} \bL_{l_j}^{-\dagger} ] \bSig_{l_j},  \ \ \forall l_j$
\State // \emph{reverse network optimization: transmit filter update}
\State \hspace{.2cm} Estimate   $\bar{\bR}_{i_k}, \bar{\bQ}_{i_k}$, and compute $\bK_{i_k}$ ,  $\forall i_k$
\State \hspace{.2cm} $\bV_{i_k} \leftarrow  \bK_{i_k}^{-\dagger}  v_{1:d} [ \bK_{i_k}^{-1} \bar{\bR}_{i_k} \bK_{i_k}^{-\dagger} ] \bLam_{i_k}, \ \ \forall i_k$
\EndFor 
\end{algorithmic}
\end{algorithm}

\subsection{Analysis and Discussions} \label{sec:discussion}
\paragraph{Interpretation}
We provide an intuitive interpretation of the problem in Lemma~\ref{lem_dlt} and its solution. 
It can be verified that $\lrb{ \alpha_i \triangleq \lambda_i[\bL^{-1} \bR \bL^{-\dagger}] }_{i=1}^r $ are also the eigenvalues of $\bQ^{-1} \bR $ (where $\bR$ and $\bQ$ represent the signal and I+N covariance matrix, respectively). Thus, $\lrb{ \alpha_i  } $ acts as a (quasi)-SINR  measure, for each of the data streams. Moreover, it can be seen that the optimal power allocated to stream $i$, $\bSig_{(i,i)}^\star$ in~\eqref{eq:optpow}, tends to zero as $\alpha_i \rightarrow 0$, i.e., no power is allocated to streams that have low-SINR.\footnote{Although the optimal power allocation to stream $i$ is zero for some streams, i.e., $\bSig_{(i,i)}=0$, in the actual implementation of the algorithm, $\bSig_{(i,i)}=\delta$ where $\delta \ll 1$.}
Moreover, note that $\lrb{\beta_i}$ represents the cost of allocating power to each of the streams (this can be seen in $(P5)$). Thus, the non-homogeneous waterfilling solution in~\eqref{eq:nhwf} allocates power to each of the streams, based on the SINR and cost of each (possibly not allocating power to some streams). 
In addition, we note that this solution reduces to that of the GMRQ problem,~\eqref{gmrqsol}, when equal power allocation is assumed.  

\paragraph{Discussion}
We now discuss the reason for adopting the equality power constraints for the problem at hand (i.e., $(J1)$ and $(J2)$), by showing the limitation of using an inequality constraint. Note that in the noise-limited regime, $\sigma_{l} \gg 1, \ \forall l \in \calL $, and consequently $\alpha_i \triangleq \lambda_i[\bL^{-1} \bR \bL^{-\dagger}]  \rightarrow 0 , \ \forall i \in \lrb{r} $. Using the fact that $\log(1+y) \approx y ,  \ y \ll 1 $, the optimal power allocation in $(P5)$ is approximated as,
\begin{align}
\sum_{i=1}^r x_i - \log_2( 1 + \alpha_i x_i ) &\approx \sum_{i=1}^r x_i -  \alpha_i x_i \nonumber  \\ 
 ~~&=  \sum_{i=1}^r (1 - \alpha_i ) x_i  \overset{ \alpha_i \rightarrow 0 }{\approx}   \sum_{i=1}^r x_i
\end{align}
When inequality constraints are considered, $(P5)$ takes the following form,  
\begin{align}
  \min \sum_i x_i \ \ \st \sum_{i=1}^r \beta_i x_i \leq \zeta, \ x_i \geq 0. 
\end{align}
One can see that the optimal solution is $x_i^\star = 0, \forall i $, and the optimal transmit/receive filter in $(J1)$ and $(J2)$ is zero. 
Thus, operating with an inequality power constraint leads to degenerate solutions, in the noise-limited regime. 
Though it might seem that an equality power constraint makes $(J1)$ and $(J2)$ harder to solve, this is not the case as both have non-convex cost functions already: despite their non-convexity, they can be effectively solved using Lemma~\ref{lem_dlt} and Lemma~\ref{lem:optpow_alloc}. 
Moreover, the convergence of BCD is unaffected since the globally optimal solution is found for each subproblem (formalized in the following proposition).  

\paragraph{Convergence of max-DLT}
Regarding convergence of the proposed algorithm, max-DLT, it is established using standard BCD convergence results. 
\begin{corollary} \label{prop:conv}
Let $\psi_n \triangleq R_{\Sigma}^{(LB)}( \lrb{\bU_{l_j}^{(n)}}, \lrb{ \bV_{l_j}^{(n)} }) , \ n = 1,2,... $ be the sequence of iterates for the objective value. Then, $\lrb{\psi_n}$ is non-decreasing in $n$, and converges to a stationary point of $R_{\Sigma}^{(LB)}( \lrb{\bU_{l_j}}, \lrb{ \bV_{l_j} })$. 
\end{corollary}
\begin{IEEEproof} $\psi_n$ converges to a stationary point of the objective, since a unique minimizer is found at each step. 
 This follows form BCD convergence results in~\cite{Tseng_convBCD_01} and~\cite[Chap 7.8]{Luenberger_LNLP_63}. 
 \end{IEEEproof}

\paragraph{Relation to other methods}
The view that the proposed approach seems close to other heuristics such as successive convex programming (SCP) and the convex-concave procedure (CCP), is slightly misleading.
Those methods start with expressions such as~\eqref{eq:DLL_lb} in Proposition~\ref{prop:dlt_bound}, then approximate $\log_2 |  \bU_{l_j}^\dagger \bQ_{l_j} \bU_{l_j} | $ with a linear function (in the case of CCP~\cite{Lipp_CCP_15}), or optimize a quadratic lower bound (in the case of SCP~\cite{Schittkowski_SCP_95}). The approximation is iteratively updated until convergence. 
Starting with~\eqref{eq:DLL_lb}, CCP~\cite{Lipp_CCP_15} generates a sequence of iterates $\lrb{\bU_{l_j}^{(n)}}_n$, where at iteration $n$, $\log_2 |  \bU_{l_j}^\dagger \bQ_{l_j} \bU_{l_j} | $ is approximated along its gradient, $\bA_{l_j}^{(n)}$, around the point $\bU_{l_j}^{(n)}$, i.e.,   
 \begin{align*} 
  \bU_{l_j}^{(n+1)} = \underset{ \bU_{l_j}  }{\argmin} \  \tr ( (\bA_{l_j}^{(n)})^\dagger \bU_{l_j} ) -  \log_2 | \bI_d + \bU_{l_j}^\dagger \bR_{l_j} \bU_{l_j} |   
 \end{align*}
Note that a comparison between the CCP updates above, and those resulting from our proposed approach, e.g., $(J1)$ and $(J2)$, indeed reveals that they different. 
Moreover, such approaches will inevitably lead to significantly larger communication overhead and complexity; this goes against the main motivation of the work (this is further discussed in Sec.~\ref{sec:overhead}).   
Thus, iteratively updating the DLT bound around the operating point (in a similar fashion to CCP or SCP), is not applicable: this is not of interest in this work, as the resulting bound would not be \emph{separable} and decouple in the transmit/receive filters. 
That being said, we will benchmark against a CCP scheme, where transmit covariance optimization was considered in the MIMO IMAC setting~\cite{Nguyen_WSR_14}.

\section{Practical Aspects}
\subsection{Comparison}
A few remarks are in order at this stage, regarding the similarities and differences between AIMS and max-DLT. Referring to the optimal update equations for each algorithm, i.e.,~\eqref{upt} and~\eqref{eq:upt2}, we clearly see that both span the same subspace, i.e. the generalized eigenspace between the signal and I+N covariance matrices. In addition, max-DLT computes the optimal power allocation for each stream. Despite this significant similarity among the two solutions, recall that they are derived from two fundamentally different separability metrics. While AIMS is an extension of max-SINR, that  greedily maximizes the separability at each BS/MS, the updates of max-DLT maximize a lower bound on the sum-rate capacity (and are shown to converge to a stationary point of the DLT bound). That being said, their performance evaluation is done via numerical results. 

\subsection{Benchmarks} \label{sec:benchmark}
As mentioned earlier, we will also investigate the proposed approach in alternate scenarios such as MIMO IBC, and the MIMO Interference Channel (MIMO IFC).
We benchmark our algorithms against widely adopted ones,
\begin{itemize}
\item[o] \emph{max-SINR}~\cite{gomadam_distributed_2011} in the MIMO IMAC, MIMO IFC and MIMO IBC
\item[o] \emph{MMSE and Weighted-MMSE}~\cite{peters_cooperative_2011, shi_wmmse_2011} in the MIMO IFC and MIMO IBC 
\end{itemize}
as well as relevant fast-converging algorithms,
\begin{itemize}
\item[o] \emph{CCP-WMMSE}~\cite{Nguyen_WSR_14}: an accelerated version of WMMSE algorithm (using CCP), for the MIMO IMAC 
\item[o] \emph{IWU}~\cite{Ghauch_IWU_15}: a fast-convergent leakage minimization algorithm for the MIMO IFC
\item[o] \emph{Uncoordinated}, wherein each transmit-receive pair perform optimal eigen-beamforming using the left/right eigenvectors of the channel (irrespective of all other pairs). 
\end{itemize}
Both IWU and CCP-WMMSE rely on the use of \emph{turbo iterations}: $I$ inner-loop iterations are carried out within each main F-B iteration, to solve a given optimization problem (analogous to \emph{primal-dual decompositions}, where an inner problem is solved to optimality, and the solution plugged into the main problem). 
While those turbo iterations are performed at the BS/MS in the case of IWU, they are run over-the-air in the case of CCP-WMMSE, thus leading to higher overhead.  
We note as well that earlier works applied SCA to MIMO IBC settings, e.g.,~\cite{Kaleva_SCPWSR_12}, but their algorithms are restricted to single-stream beamforming/combining.

\subsection{Distributed CSI Acquisition} \label{sec:dist_csi_acqusition}
We underline in this section some practical issues that relate to the proposed schemes, such as the mechanism for distributed CSI acquisition, and the resulting communication overhead and computational complexity. Although additional issues such as robustness have to considered as well, such matters are outside the scope of the current paper. 
We reiterate the fact that CSI acquisition mechanisms are outside the scope of the paper (we refer the reader to~\cite{Brandt_DistCSI_15}). We just summarize the basic operation behind F-B iterations.  

Evidently, the operation of such schemes is contingent upon each transmitter / receiver  being able to estimate the signal and the I+N covariance matrices, in a fully distributed manner. From the perspective of this work, this is accomplished via the use of \emph{precoded pilots} to estimate the \emph{effective channels}.\footnote{A full investigation of the total overhead of this decentralized solution, as compared to a centralized implementation, falls outside the scope of the current paper.}
In the forward phase, the signal covariance matrix, as well as the I+N covariance matrices, can be computed after estimating the effective signal channel, and the effective interfering channels, respectively. The receive filters at the base stations are updated following any of the proposed algorithms (summarized in Fig.~\ref{fig:frame}). Then, in the downlink phase, the same procedure is used to estimate the signal and I+N covariance matrices, and update the filters at the receivers. This constitutes one F-B iteration. Recall that $T$ is the total number of such iterations, that are carried out. 

\subsection{Communication Overhead} \label{sec:overhead}
For such schemes to be fully distributed, the required CSI quantities have to be obtained via uplink-downlink pilots. Each F-B iteration has an associated communication overhead, namely that of bi-directional transmission of pilots. We adopt a simplistic definition of the communication overhead, as the minimal number of orthogonal pilots symbols, required for estimating the required CSI quantities (keeping in mind that the actual overhead will be dominated by this quantity). We note that almost all prior algorithms that have been proposed in the context of cellular systems, focus on a regime with a high enough number of F-B iterations, i.e., $T=100 \sim 1000$, such as~\cite{gomadam_distributed_2011,schmidt_minimum_2009,peters_cooperative_2011,shi_wmmse_2011} to list a few. In addition, the relatively larger number of antennas in lower bands of mmWave systems, results in significantly more iterations. 
 On the contrary, and in line with recent attempts such as~\cite{Komulainen_EffCSI_13, Brandt_FastConv_15, Ghauch_IWU_15}, we assume that this modus operandi is not feasible in the systems we consider (since F-B iterations are carried out over-the-air, and the associated overhead would be higher than the potential gains). Indeed the lower coherence time of mmWave channels (compared to that of traditional systems) makes the low-overhead requirement even more stringent. 
 We thus focus on a regime where $T = 2 \sim 5$.   
In addition, we assume that the minimal number of orthogonal pilots is used, i.e., $d$ orthogonal pilot slots for each uplink/downlink effective channel. Moreover, the pilots are orthogonal across users and cells, resulting in a total of $KLd$ orthogonal pilots for each uplink/downlink training phase. Consequently, the total overhead of both AIMS and max-DLT is approximately,  
\begin{align*}
\Omega_{\textrm{prop}} = T( \underbrace{KLd}_{  UL } + \underbrace{KLd}_{ DL } ) = 2TKLd  \ \textup{channel uses} .
\end{align*}
It can be verified that the overhead is the same for benchmarks such as max-SINR, IWU and MMSE. 
Moreover, using similar arguments one can approximate the overhead of CCP-WMMSE and WMMSE (in the number of channel uses), as      
\begin{align*}
&\Omega_{\textrm{ccp-wmmse}} = T[(\underbrace{KLM}_{ \substack{UL \ chann. \\ estim } }) \underbrace{\times (L-1)}_{ \substack{CSI \\ sharing}} +  \underbrace{I}_{turbo} \times ( \underbrace{KLN}_{\substack{cov. \ mat \\ upd.}}) ]    \\ 
&\Omega_{\textrm{wwmse}} = T( \underbrace{KLd}_{ UL} + \underbrace{KLM}_{ weights} + \underbrace{KLd}_{ DL})    
\end{align*}
where $I$ denotes the number of turbo iterations. Though a coarse measure, we can see that the overhead associated with WMMSE and its fast-converging variant CCP-WMMSE are significantly higher than that of the proposed schemes. Moreover, CCP-WMMSE exhibits massively larger overhead than the other two, due to the fact that the turbo  optimization is carried over-the-air (as described in Sec.\ref{sec:benchmark}), and that the CSI for the uplink channels is shared among the BSs~\cite{Nguyen_WSR_14}. The overhead of the aforementioned schemes will be included in the numerical results.

\subsection{Complexity}
Despite the fact that the communication overhead is the limiting resource in cellular networks, we nonetheless shed light on the complexity of the proposed approaches, for completeness. 
The computational complexity of both AIMS and max-DLT is dominated by the Cholesky Decomposition of the I+N covariance matrix, and the Eigenvalue Decomposition of $\bM$, both of which have similar complexity of $\mathcal{O}(N^3)$ (keeping in mind that other operations such as matrix multiplication and bisection search are quite negligible in comparison). Thus, the complexity (per F-B iteration) is approximately,
\begin{align*}
C_{\textup{prop}} = \calO(2KL(M^3 + N^3) ) \ .
\end{align*}
Note that the same holds for benchmarks such as max-SINR, IWU, and WMMSE since they are dominated by matrix inversion of the I+N covariance matrix. 
While the complexity of CCP-WMMSE is also dominated by the above quantity, it also involves running a series of semi-definite programs (using interior point solvers), within each turbo iteration. This renders the algorithm quite costly.

\section{Numerical Results}
\subsection{Simulation Methodology}
We use the achievable sum-rate in the network as the performance metric, where the achievable user rate is given by~\eqref{eq:user_rate}. 
Because the approach here is presented in the context of MIMO IMAC, a significant fraction of the results will be under this setup.
The model, algorithms and results presented in the paper, are applicable to the MIMO IBC: in the latter setting, the transmit filters ($\lrb{\bV_{i_k}}$) are at the BSs, receive filters ($\lrb{\bU_{l_j}}$) at the MSs, and $M$ (resp. $N$) denotes the number of BS (resp. MS) antennas. For example, a system with $32$ BS antennas and $4$ MS antennas, is modeled by setting $M=4,N=32$ in the MIMO IMAC case, and $N=4,M=32$ in the MIMO IBC.  

We limit the number of F-B iterations, $T$, to a small number. Following the calculations in Sec.\ref{sec:overhead}, we include in the results the overhead, $\Omega$, for each of the algorithms, as a function of $T$ (and the number if turbo iterations, $I$, when applicable). 
Moreover, the considered sub-$28$ GHz systems allow for more BS/MS antennas compared to traditional cellular frequencies, for the same transmit/receive array surface (due to the antenna spacing being smaller). This will be reflected in the numerical results. 
Finally, we note that all curves are averaged over $500$ channel realizations.
In the first part, we assume a block-fading channel model with static users, having i.i.d. channels, to benchmark against several known schemes, and canonical channel model. In another part, we will build a more realistic simulation setup that is more reflective of the systems under consideration.

\subsection{Part I} 
\begin{figure}
  \center
  \includegraphics[ height=6cm, width=8.5cm ]{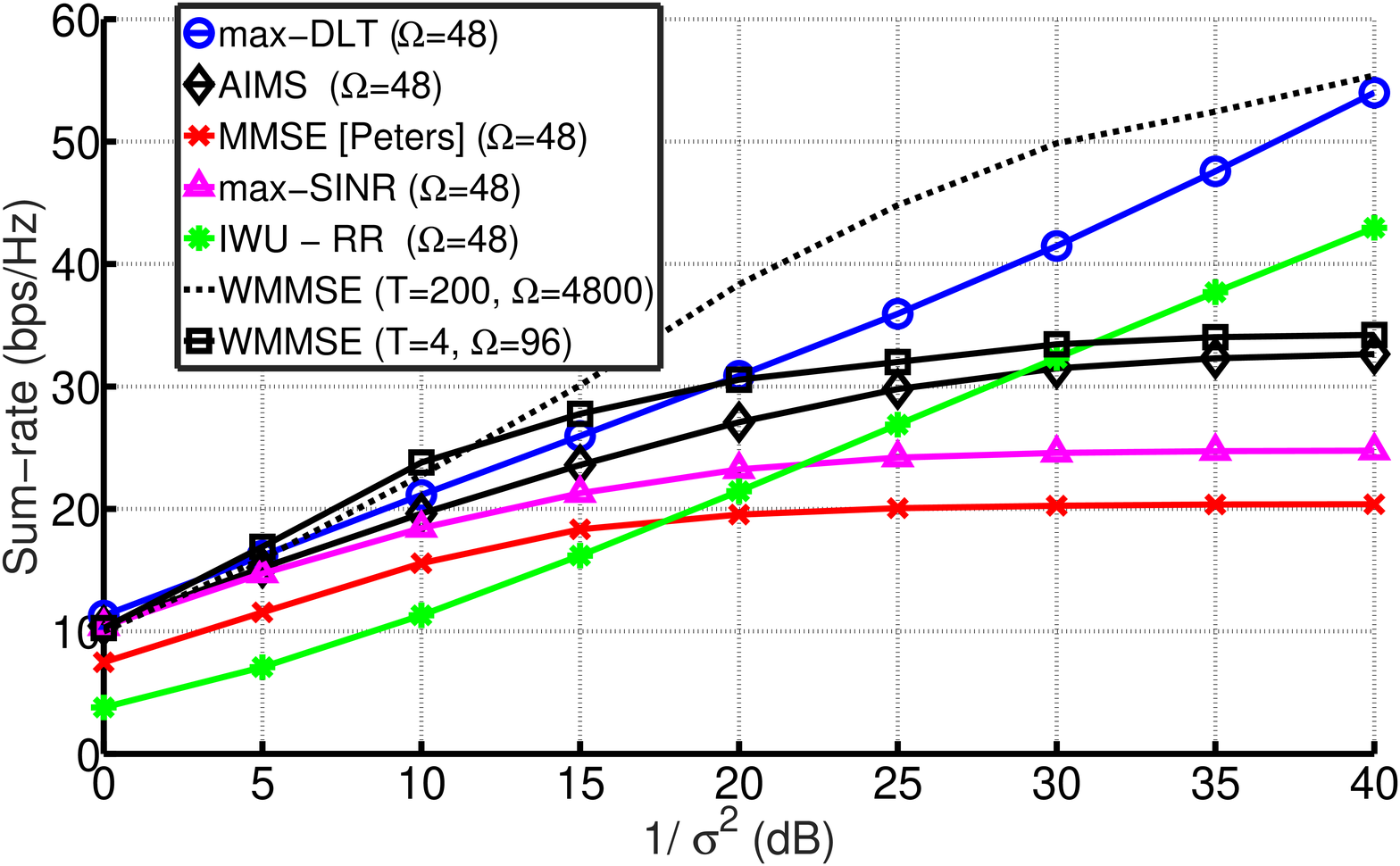}
  \caption{Ergodic sum-rate vs $1/\sigma^2$, for $L=3, K=1, M=N=4, d=2, T = 4$ (MIMO IFC)} 
  \label{fig:feasibleifc} 
\vspace{.5cm} 
\end{figure}

\paragraph*{Single-user Multi-cell MIMO Uplink}
We will first evaluate the performance of our schemes in a MIMO Interference Channel (IFC), where many schemes such as 
max-SINR~\cite{gomadam_distributed_2011}, MMSE~\cite{peters_cooperative_2011}, and IWU~\cite{Ghauch_IWU_15} are applicable. 
We set $M=N=4, d=2$ and fix the number of F-B iterations, $T=4$, for all schemes.  
We also included Weighted-MMSE with the corresponding number of F-B iterations ($T=4$), and a large enough number of iterations (as an upper bound). 
It is clear from Fig.~\ref{fig:feasibleifc} that while max-DLT has similar performance as WMMSE (for $T=4$) in the low-to-medium SNR regime, the gap increases in the high-SNR region ($\text{SNR} \geq 20 \text{ dB}$). 
Moreover, we note that our proposed schemes outperform all other the benchmarks, across all SNR regimes. 
In particular, the performance gap between max-DLT and the benchmarks, is quite significant in the medium-to-high SNR region. 
Moreover, max-DLT offers a $35\%$ gain over the fast-converging IWU: Though both are able to turn off some streams in view of reducing interference, max-DLT also optimizes the signal subspace as well. 
Note that the `optimal-performance' of WMMSE is achieved for $T=200$, but the resulting overhead is massive.  
Although the performance of max-DLT is similar to WMMSE ($T=4$) in low-to-medium SNR regime, the overhead is much lower for the former. Moreover, the gap increases in the medium-to-high SNR region.  
The rest of results will show that the performance gap between our proposed algorithms and several known benchmarks, increases in the regime of interest (low-overhead, large system dimensions). 

\begin{figure}
	\center
	\includegraphics[trim={1.5cm 0 0 1.5cm}, clip, scale=.22]{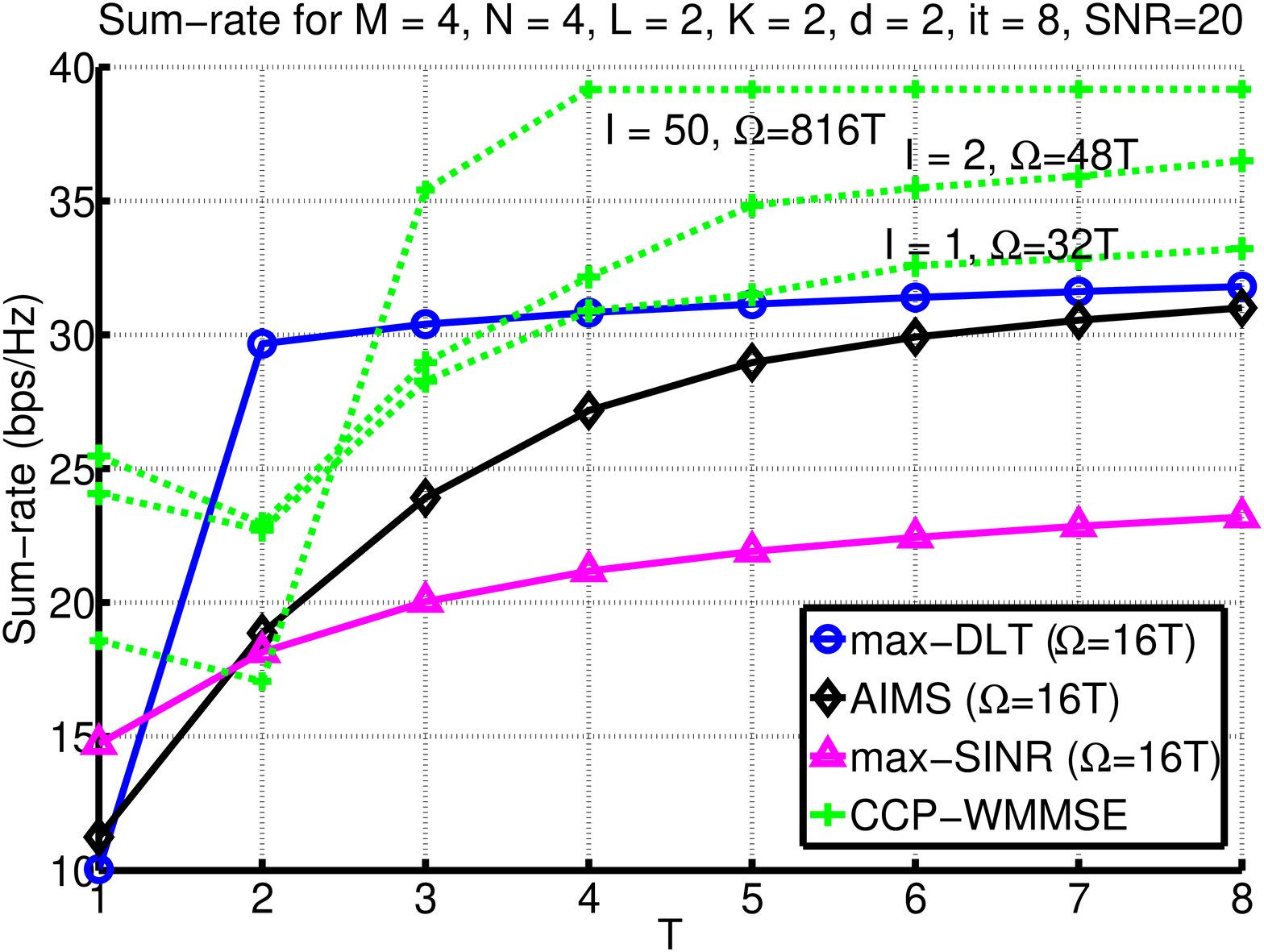}
	\caption{Ergodic sum-rate vs $T$, for $L=2, K=2, M=4, N=4, d=2$ } 
	\label{fig:SR_CCP}
\end{figure}

\paragraph*{Multi-user Multi-cell MIMO uplink}
We next evaluate a MIMO IMAC setting with  $L=2, K=2, M=4, N=4, d=2$, as a function of the number of F-B iterations, $T$. 
We also benchmark against CCP-WMMSE (summarized in Sec.~\ref{sec:benchmark}) by varying the number of turbo iterations $I$, and testing the resulting performance and overhead.  
Fig.~\ref{fig:SR_CCP} clearly shows the fast-converging features of both algorithms. More specifically, this is apparent in the case  of max-DLT, that reaches $95 \%$ of its performance in $2$ iterations. 
While the performance of max-DLT is slightly better than CCP-WWMMSE for $I=1$, the overhead of the latter is twice that of the former (CCP-WMMSE becomes better than max-DLT for $I=2$, but the  resulting overhead is thrice as high). Note that the `full' performance CCP-WMMSE is achieved for $I=50$, but the the resulting overhead (and complexity) are orders-of-magnitude larger than the proposed schemes. Its performance is quite sensitive to solving the inner problem to optimality  (i.e., until the turbo iteration converges), thus making it ill-suited for larger setups. 
Indeed, the running time of CCP-WMMSE (using the Mosek solver in CVX) prevented us from testing its performance for larger antenna configurations. 

\begin{figure}
	\center
	\includegraphics[scale=.19]{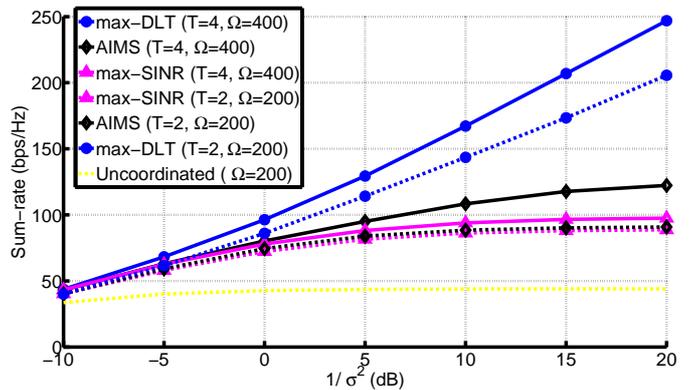}
	\caption{Ergodic sum-rate vs $1/\sigma^2$, for $L=5, K=5, M=4, N=32, d=2$ (Uplink) } 
	\label{fig:massiveUL}
\end{figure}

\paragraph*{Large-scale Multi-user Multi-cell MIMO uplink}
We test the performance of a system where a larger number of antennas is available at the BS (enabled by sub-$28$ GHz systems). We evaluate a large-scale (in the number of antennas at the BS) multi-user multi-cell uplink with $L=5, K=5, d=2, M=4, N=32$. Fig.~\ref{fig:massiveUL} shows the resulting sum-rate of the proposed schemes (and max-SINR), for $T=2$ and $T=4$ (we were unable to include CPP-WMMSE as the resulting simulation time was too long to be included). Recall that for each of the simulated values of $T$, the overhead is the same for all schemes. 
We observe that both our schemes outperform max-SINR significantly. In particular, max-DLT offer twice the performance of max-SINR at $5$dB (this performance gap increases with the SNR). 
And while both our schemes show significant performance gain by increasing $T$, the corresponding gain that max-SINR exhibits is negligible in comparison.

\begin{figure}
	\center
	\includegraphics[scale=.19]{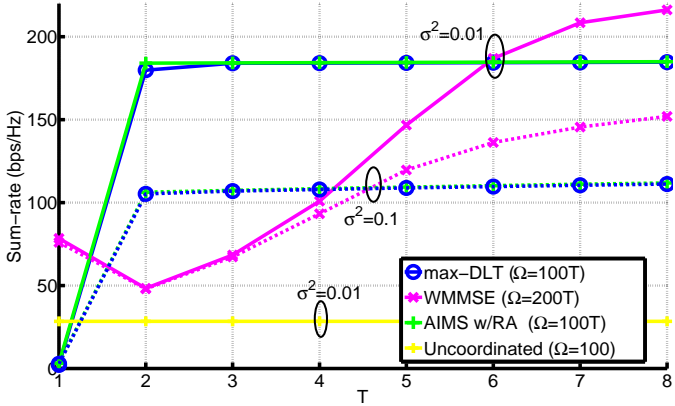}
	\caption{Ergodic sum-rate vs $T$, for $L=5, K=5, M=32, N=4, d=2$ (Downlink). Solid curves correspond to noise power $\sigma^2=10^{-2}$, and dashed ones to $\sigma^2 = 10^{-1}$.   } 
	\label{fig:massiveDL}
\end{figure}

\paragraph*{Large-scale Multi-user Multi-cell MIMO downlink} 
We next investigate a dual communication setup of the one just above, a MIMO IBC obtained by setting $M=32, N=4$ (all else being the same). 
We benchmark our results against the well-known WMMSE algorithm~\cite{shi_wmmse_2011}. 
Note that while WMMSE employs a sum-power constraint, our schemes have a per-user power constraint, and thus assume equal power allocation among the users.\footnote{If $P_t$ is the per user constraint for our schemes, then $K P_t$  is the per-BS sum-power constraint for WWMMSE.} This implies that a more stringent constraint is placed on our schemes.
Despite this unfavorable setup, both our schemes significantly outperform WMMSE, the gap becoming quite large when noise power is $\sigma^2=0.01$ (as seen in Fig.~\ref{fig:massiveDL}). Note as well that the overhead of our proposed schemes, is half that of WMMSE as the latter requires feedback of the weights (refer to Sec.~\ref{sec:overhead} for the overhead calculations). Needless to say, the full-performance that WMMSE is expected to deliver, is reached after more iterations are performed. 
The reason behind this behavior is the fast-converging nature of our algorithms, allowing them to reach a good operating point, in just $2$ iterations. In the case of the max-DLT, this is due to the stream control feature of the non-homogeneous waterfilling. 

While Figs.~\ref{fig:massiveUL}-\ref{fig:massiveDL} have the same configuration (same number of BS/MS antennas, data streams, users and cells), as far as the algorithms are concerned, $M$ and $N$ are different ($N=32, M=4$ in the uplink, and $N=4, M=32$ in the downlink). Consequently, comparing the performance of any scheme in both configurations, is not informative. 
This is further reinforced by the absence of any result linking sum-rate, for the MIMO IMAC and the MIMO IBC (unlike sum-rate in the MIMO MAC and the MIMO BC, related by duality).

\subsection{Part II: }
In this part we use a more comprehensive simulation setup.
For lack of channel measurements in the $6-18$ GHz band, we will use recent results in the $28$ GHz band~\cite{MacCartney_PLmmW_14}, in the non line-of-sight setting. 
The MIMO channel $\bH_{l,i_k}$, from user $i_k \in \calI$ to BS $l \in \calL$, has coefficients
\begin{align} \label{eq:chann}
[\bH_{l,i_k}]_{p,q} = \sqrt{\gamma_{l,i_k}} g_{p,q} , \ \ \forall (p,q) \in \lrb{N}\times \lrb{M} .
\end{align}
In the above, $ (\gamma_{l,i_k})_{dB} =  20\log_{10}(4 \pi/\lambda_c) + 10 n_p \log_{10}(D_{l,i_k}) + \psi_{l,i_k}$ is the pathloss between user $i_k \in \calI$ and BS $l \in \calL$, where $D_{l,i_k}$ is the corresponding distance, $n_p =3.4$ the pathloss exponent, $\lambda_c$ the carrier wavelength (corresponding to $28$ GHz), and $\psi_{l,i_k}$ is log-normal with zero mean and variance of $9$dB~\cite{Rappaprt_mmWprop}. Moreover, $g_{p,q}$ follows a Rician distribution with zero mean and unit variance, to model the line-of-sight components. 
We consider a `dense' multi-user multi-cell setup with $L=9$ cells, each with radius $10$m and serving $K=8$ users (dropped uniformly within the cell). We investigate both uplink and downlink communication.  

\begin{figure}
	\center
	\includegraphics[trim={0 0 0 1.5cm}, clip, scale=.19]{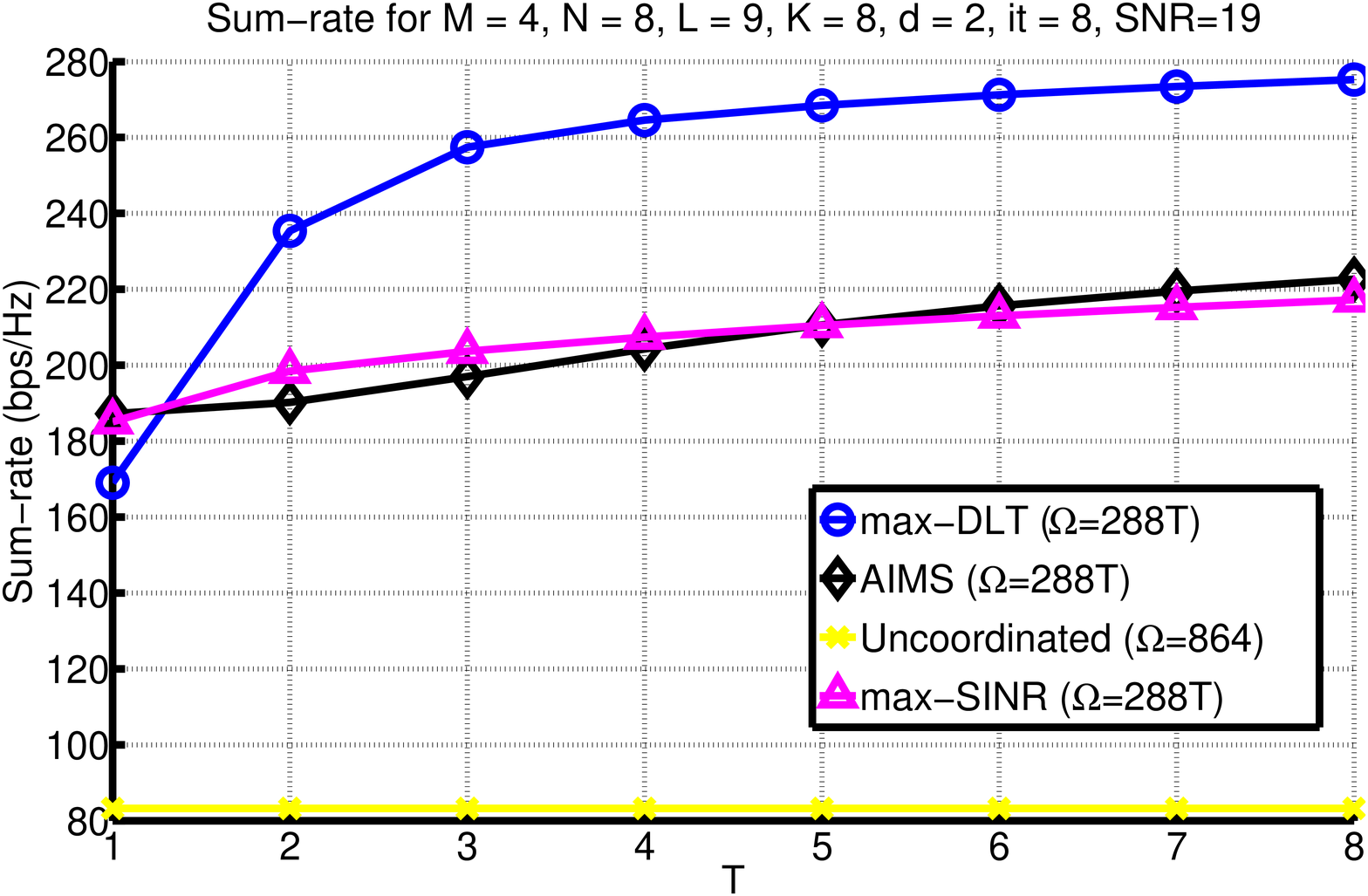}
	\caption{Ergodic sum-rate vs $T$, for dense uplink ($N=8, M=4, d=2, L=9, K=8$).  } 
	\label{fig:denseUL}
\end{figure}

\begin{figure}
	\center
	\includegraphics[trim={0 0 0 1.5cm}, clip, scale=.2]{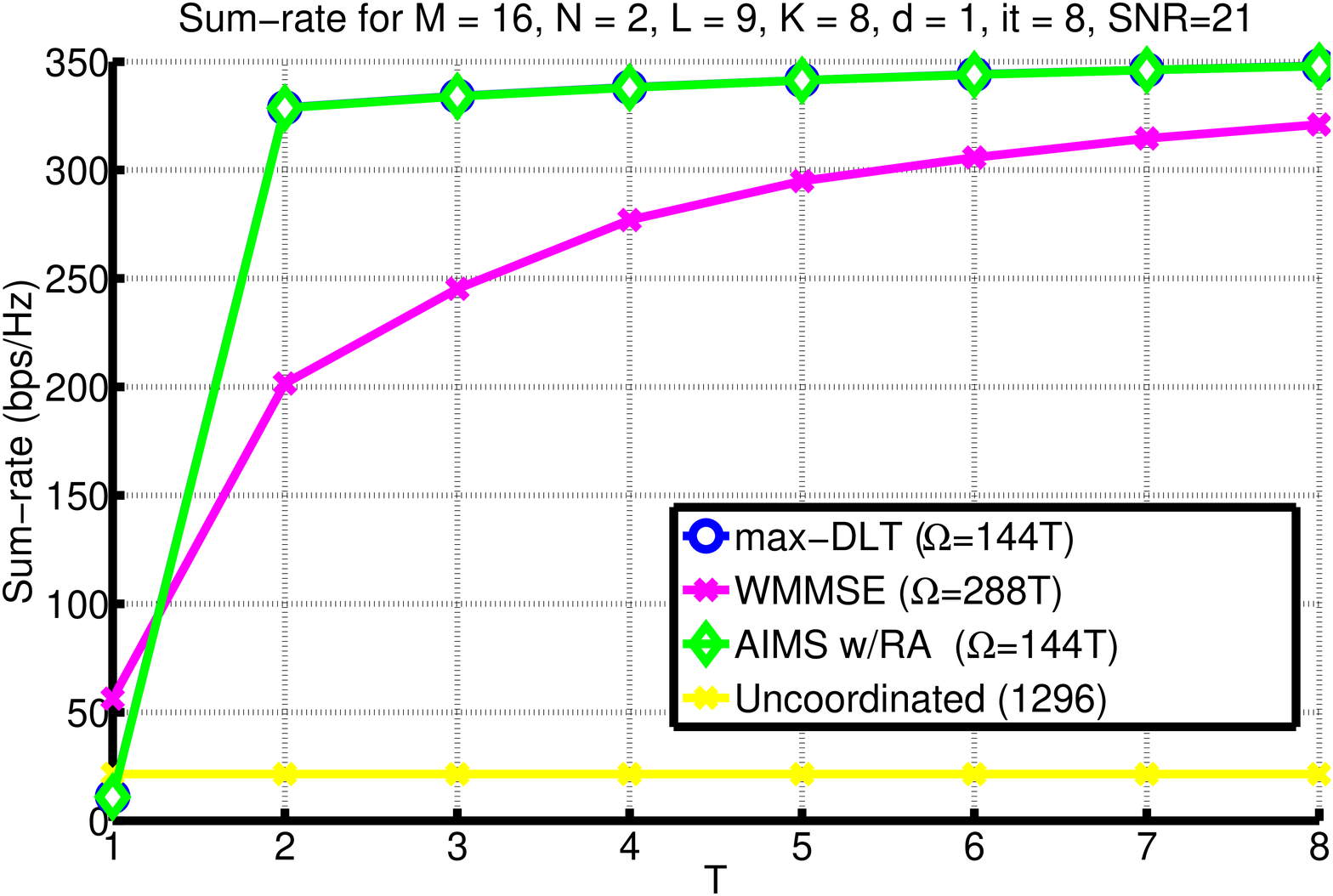}
	\caption{Ergodic sum-rate vs $T$ for dense downlink ($M=16, N=4, d=1, L=9, K=8$)  } 
	\label{fig:denseDL}
\end{figure}

\paragraph*{Dense Multi-user Multi-cell uplink}
BSs are equipped with $N=8$ antennas, and MSs with $M=4$, sending $d=2$ streams each. 
The ergodic sum-rage is shown in Fig.~\ref{fig:denseDL}, as a function of $T$, and the average effective SNR (including pathloss) is around $19$dB. 
As we can see in Fig.~\ref{fig:denseUL}, max-DLT provides significant gains over the other schemes. 
Moreover, the poor performance of the uncoordinated scheme further confirms our motivation for this work: mmWave systems in the lower bands, are not selective enough to bypass the need for interference management; indeed \emph{any} of the coordination schemes considered here doubles the sum-rate performance (Fig.\ref{fig:denseUL}).   
Note that the overhead for max-DLT at $T=3$, is similar to that of the uncoordinated transmission, while providing around three times higher sum-rate. 

\paragraph*{Dense Multi-user Multi-cell downlink}
We also investigate another `dense' setting ($M=16, N=4, d=1, L=9, K=8$), with downlink communication (using the same simulation setup). 
The power constraint on WMMSE follows the same convention as the large-scale multi-cell downlink above, and the observed trends are still the same (as seen in Fig.~\ref{fig:denseDL}). Despite the unfavourable setup, our schemes offer a massive performance gain over WMMSE, with half the resulting overhead (the curves for max-DLT and AIMS w/RA are overlapping). We reiterate the fact that this is due to their ability to shut down streams with low-SINR, and thus converging to good sum-rate points, in less than $2$ iterations. 

While max-DLT has the distinct feature of quickly converging (due to the built-in stream-control), WMMSE has a slower convergence converge to a stationary point of the sum-rate. In this work, we leverage this feature of max-DLT to significantly reduce the overhead (refer to corresponding $\Omega$ in Figs.~\ref{fig:massiveDL},\ref{fig:denseDL}). 
However, in general, the performance of WMMSE will exceed that of max-DLT, as T increases: this occurs at $T=6$ iterations in the case of Fig.~\ref{fig:massiveDL}, and $T>8$ in Fig.~\ref{fig:denseDL}).

\subsection{Discussions}
As mentioned earlier, the non-homogeneous waterfilling solution clearly shows that streams that have low SINR are turned-off, and power is only allocated to the ones that exhibit relatively high SINR. This greatly speeds up the convergence of max-DLT, and allows it to achieve its required performance, with that limited number of F-B iterations (e.g., $2$). On the other hand, due to the large dimensions inherent to lower band mmWave systems (i.e., more antennas, cells, users) other benchmarks will require more iterations to reach a similar performance.
In addition, the rank-adaptation feature in AIMS offers a trade-off between reducing interference and diversity of the signal: however, in interference-limited scenarios, the former is more critical than the latter. The improved performance from RA, shown in the numerical results section, validates this proposition. 
As for the overhead, our schemes are based on the framework of F-B iterations and result in minimal overhead (the overhead consisting of uplink/downlink pilots only). However, other schemes such as WMMSE and CCP-WMMSE require additional pilots and feedback, and result in significantly higher overhead (as detailed in Sec.~\ref{sec:overhead}). Moreover, the evaluations from the realistic dense uplink and downlink setup both conclude that interference management is a vital component in sub-$28$ GHz mmWave systems. 

\section{Conclusion}
In this work, we shed light on the need for interference management in lower bands of the mmWave spectrum, while highlighting the inapplicability of conventional approaches for distributed coordination.
We thus proposed AIMS, a distributed algorithm that alternately maximizes the separability metric, for both the uplink and downlink, and established the fact that this is a generalization of the well-known max-SINR. Moreover, we advocated the use of DLT bounds, and highlighted their significant advantage in yielding optimization problems that decouple at both the transmitters and receivers. We provided a generic solution to this problem, the so-called non-homogeneous waterfilling (underlining its built-in stream-control feature), and proposed another distributed algorithm, max-DLT, that solves the problem in a distributed manner. Convergence to a stationary point of the DLT bound was also established.  
We later verified through extensive simulations that our proposed algorithms significantly outperform other well-known schemes, in the desired low-overhead regime (while still requiring less overhead). 
Moreover, the results also confirmed the need for interference management, and that the proposed approaches are a good fit for the this task. 
 

\appendix



\subsection{Sketch of proof for Proposition~\ref{prop:nonunitary}} \label{A3a}
We need to show that $ (\bU_{l_j}^\star)^\dagger \bU_{l_j}^\star  \approx \alpha \bI_{d} $ happens with probability zero.
Note that due to the i.i.d. nature, of MIMO channel coefficients, the eigenvalues of $\bQ_{l_j}$ can be assumed to be distinct (and $\bQ_{l_j}$ is full rank), almost surely. Then, it can be verified that the same holds for $\bL_{l_j}, \bL_{l_j}^\dagger  $  and $ (\bL_{l_j}^{\dagger} \bL_{l_j})^{-1}$. Then, $(\bL_{l_j}^{\dagger} \bL_{l_j})^{-1} \approx \alpha \bI_N$ happens with probability zero. 
Recalling that $\bPsi_{l_j}^\dagger \bPsi_{l_j}  = \bI_{d} $, then the following equivalent statements happen with probability zero,
\begin{align*}
&\bPsi_{l_j}^\dagger   (\bL_{l_j}^{\dagger} \bL_{l_j})^{-1}  \bPsi_{l_j} \approx    \bPsi_{l_j}^\dagger  ( \alpha \bI_N )  \bPsi_{l_j}   \\
&\Leftrightarrow ( \bPsi_{l_j}^\dagger   \bL_{l_j}^{-1} ) (\bL_{l_j}^{-\dagger} \bPsi_{l_j}) \approx \alpha \bI_d \Leftrightarrow (\bU_{l_j}^\star)^\dagger \bU_{l_j}^\star  \approx \alpha \bI_{d} 
\end{align*}

\subsection{Proof of Proposition~\ref{prop:dlt_bound} } \label{app:dlt_bound}
We start by lower bounding the user rate in~\eqref{eq:user_rate}, as
\begin{align}
r_{l_j} & \geq \log_2 | (\bU_{l_j}^\dagger \bQ_{l_j} \bU_{l_j})^{-1} + (\bU_{l_j}^\dagger \bR_{l_j} \bU_{l_j})(\bU_{l_j}^\dagger \bQ_{l_j} \bU_{l_j})^{-1} | \nonumber \\ 
&= \log_2 |(\bI_d + \bU_{l_j}^\dagger \bR_{l_j} \bU_{l_j})(\bU_{l_j}^\dagger \bQ_{l_j} \bU_{l_j})^{-1} |  \nonumber \\
&= \log_2 | \bI_d + \bU_{l_j}^\dagger \bR_{l_j} \bU_{l_j} |  - \log_2 |  \bU_{l_j}^\dagger \bQ_{l_j} \bU_{l_j} |  \nonumber \\ 
&\geq \log_2 | \pmb{I}_d +  \bU_{l_j}^\dagger \bR_{l_j} \bU_{l_j} |  - \tr(  \bU_{l_j}^\dagger \bQ_{l_j} \bU_{l_j} ) \triangleq r_{l_j}^{(LB)} \label{eq:trlb}
\end{align}
where the first inequality follows from combining A2) in~\eqref{eq:intf_limited}, and the monotonically increasing nature of $\log|\bX|$. Moreover the last one follows from using $\log|\bA| \leq \tr(\bA)$ for $\bA \succeq \pmb{0} $. We rewrite $r_{l_j}$ in~\eqref{eq:user_rate} as,
\begin{align*}
r_{l_j} &= \log_2 | (\bU_{l_j}^\dagger \bR_{l_j} \bU_{l_j})(\bU_{l_j}^\dagger \bQ_{l_j} \bU_{l_j})^{-1} [ \bI_d  \\
&~~~+ (\bU_{l_j}^\dagger \bQ_{l_j} \bU_{l_j})(\bU_{l_j}^\dagger \bR_{l_j} \bU_{l_j})^{-1}   ]  |  \\
&= \log_2 | (\bU_{l_j}^\dagger \bR_{l_j} \bU_{l_j}) (\bU_{l_j}^\dagger \bQ_{l_j} \bU_{l_j})^{-1} |  \\
&~~~+ \log_2| \bI_d + (\bU_{l_j}^\dagger \bQ_{l_j} \bU_{l_j})(\bU_{l_j}^\dagger \bR_{l_j} \bU_{l_j})^{-1}     |  \\
&=\log_2 | \bU_{l_j}^\dagger \bR_{l_j} \bU_{l_j}  |  - \log_2 |\bU_{l_j}^\dagger \bQ_{l_j} \bU_{l_j} |  \\
&~~~+ \calO(\tr[(\bU_{l_j}^\dagger \bQ_{l_j} \bU_{l_j})(\bU_{l_j}^\dagger \bR_{l_j} \bU_{l_j})^{-1}  ])
\end{align*}
Thus, $r_{l_j}$ is approximated by $\log_2 | \bU_{l_j}^\dagger \bR_{l_j} \bU_{l_j}  |  - \log_2 |\bU_{l_j}^\dagger \bQ_{l_j} \bU_{l_j} |$ (where the error is given in the above equation). Plugging this result in $\Delta_{l_j}$ yields, 
\begin{align*}
\Delta_{l_j} &=\log_2|\bU_{l_j}^\dagger \bR_{l_j} \bU_{l_j}| - \log_2|\bU_{l_j}^\dagger \bQ_{l_j} \bU_{l_j}| \\
&~~~- [\log_2 | \pmb{I}_d +  \bU_{l_j}^\dagger \bR_{l_j} \bU_{l_j} |  - \tr(  \bU_{l_j}^\dagger \bQ_{l_j} \bU_{l_j} )] \\
&~~~+ \calO(\tr[(\bU_{l_j}^\dagger \bQ_{l_j} \bU_{l_j})(\bU_{l_j}^\dagger \bR_{l_j} \bU_{l_j})^{-1}  ]) 
\end{align*}
Referring to the above, in the interference-limited regime, i.e., A1) in~\eqref{eq:intf_limited}, the first and third terms become negligible w.r.t. the second and fourth. Consequently, 
\begin{align*}
\Delta_{l_j} &= \tr(  \bU_{l_j}^\dagger \bQ_{l_j} \bU_{l_j} )  -  \log_2|\bU_{l_j}^\dagger \bQ_{l_j} \bU_{l_j}|   \\
&~~~+  \calO(\tr[(\bU_{l_j}^\dagger \bQ_{l_j} \bU_{l_j})(\bU_{l_j}^\dagger \bR_{l_j} \bU_{l_j})^{-1}  ])
\end{align*}

\subsection{Proof of Lemma~\ref{lem_dlt} } \label{A1} 
We rewrite the problem into a series of equivalent forms.
Letting $ \bZ = \bL^{\dagger} \bX \Leftrightarrow \bX = \bL^{-\dagger} \bZ $, then $(P)$ in ~\eqref{opt:logtr} is equivalent to, 
\begin{align*}
(P2) \ \begin{cases}  
     \underset{\bZ}{\min} \ f(\bZ) \triangleq  \tr( \bZ^\dagger \bZ ) - \log_2 | \pmb{I}_d + \bZ^\dagger \bM \bZ |   \\
     \st \ \tr(\bZ^\dagger \bA \bZ)=\zeta  \\
\end{cases}
\end{align*}
where $\bA = (\bL^\dagger \bL)^{-1}$. Letting $\bZ = \bT \bSig \bV^\dagger$ be the SVD of $\bZ$ ($\bT \in \mathbb{C}^{n \times r}, \bSig \in \mathbb{R}^{r \times r}$) we rewrite $(P2)$ into an equivalent form, 
\begin{align*}
(P3) \  
\begin{cases}  
    \underset{\bT , \bSig }{\min} \  \tr( \bSig^2 ) - \log_2 | \pmb{I}_d + \bSig^2 \bT^\dagger \bM \bT |  \\
    \st \ \tr( \bSig^2 \bT^\dagger \bA \bT)=\zeta  \\
\end{cases}
\end{align*}

Let us first look only at the objective in $(P3)$, to illustrate the argument. 
Note that for any given $\bSig$, the optimal $\bT$ is given by $\bT^\star \triangleq v_{1:r}[\bM] = \bPsi $. 
\footnote{This follows from maximizing $\log_2 |  \pmb{I}_d +  \bT^\dagger \bM \bT | $, over the set of unitary matrices $\bT$. Recall that  $ \bU^\star = \argmax_{\bU^\dagger \bU= \bI } ~ |\bI + \bU^\dagger \bS \bU| =\argmax_{\bU^\dagger \bU= \bI } ~ |\bU^\dagger \bS \bU|  ~,  ~\bS \succeq \pmb{0} $. Then, it is well known that $\bU^\star \triangleq v_{1:r}[\bS] $ \cite{golub_matcomp_96}.}   
Moreover, $\bT^\star$ does not depends on $\bSig$ (only on $\bM$): thus, $\bT^\star$ can be plugged into the objective, and one can solve for $\bSig$. Though the presence of a constraint makes this approach not suitable in general, note that this particular constraint allows scaling of the optimal solution, to always satisfy the constraint (this becomes clear when we express the problem as a function of the columns of $\bT$, and the diagonal entries in $\bSig$). This can also be checked by considering solutions of the form $\bT \neq v_{1:r}[\bM]$, and showing that they cannot be optimal. 
With that in mind, the feasible set of $(P3)$ becomes $\tr( \bSig^2 \bPsi^\dagger \bA \bPsi ) = \sum_i \sigma_i^2 \beta_i  $, where $\lrb{\sigma_i}$ are the diagonal elements of $\bSig$. Thus, $(P3)$ becomes, 
\begin{align*}
(P4) \begin{cases}
 \underset{ \lrb{\sigma_i} }{\min} \  \sum_{i=1}^r \left( \sigma_i^2 - \log_2(1+\alpha_i \sigma_i^2 ) \right) \\  
 \st \ \sum_{i=1}^r \beta_i \sigma_i^2  = \zeta  
 \end{cases}
\end{align*} 
Letting $x_i = \sigma_i^2 $, we can rewrite the problem as,
\begin{align*}
(P5)
\begin{cases}
\underset{ \lrb{x_i} }{\min} \sum_{i=1}^r \left( x_i - \log_2( x_i + \frac{1}{\alpha_i}) \right) \\ 
\st \ \sum_{i=1}^r \beta_i x_i = \zeta, \ x_i \geq 0, \forall i  
\end{cases}
\end{align*} 
$(P5)$ is a generalization of the well-known waterfilling problem: in fact, $(P5)$ reduces to the waterfilling problem, if $\beta_i = 1, \forall i$, and by dropping the first term in the objective. We start by writing the associated KKT conditions. 
\begin{align*}
\begin{cases}
1 - (x_i + \alpha_i^{-1})^{-1}  + \mu \beta_i - \lambda_i = 0, \ \forall i \\
\sum_i \beta_i x_i = \zeta , \ x_i \geq 0   \\
\lambda_i x_i = 0, \ \lambda_i \geq 0,  \ \mu \neq 0, \forall i
\end{cases}
\end{align*}
Firstly, note that $\lambda_i$ act as slack variables and can thus easily be eliminated. Considering two cases, $\lambda_i = 0, \forall i$ or $\lambda_i > 0, \forall i$, the optimal solution can be easily found as, 
\begin{align*}
x_i^\star &= 
\begin{cases}
(1+\mu \beta_i)^{-1} - \alpha_i^{-1}, \ \textrm{if} \ \mu < (\alpha_i - 1)/\beta_i  \\
0, \ \textrm{if} \ \mu > (\alpha_i - 1)/\beta_i  \\
\end{cases} \\
&= \Big( 1/(1+\mu^\star \beta_i) - 1/\alpha_i  \Big)^+ , \forall i 
\end{align*}
where $\mu^\star$ is the unique root to  
\begin{align*}
g(\mu) \triangleq \sum_{i=1}^r \beta_i \Big( 1/(1+\mu \beta_i) - 1/\alpha_i  \Big)^+  - \zeta
\end{align*}
Note that $g(\mu)$ is monotonically decreasing, for $\mu > -1/(\max_i \beta_i )  $, and $\mu^\star$ can be found using standard 1D search methods, such as bisection. Thus, the optimal solution for $(J1)$ is $\bZ^\star = \bPsi \bSig^\star $(where $\bSig^\star_{(i,i)} =  \sqrt{x_i} , \forall i $), and that of~\eqref{opt:logtr} is $\bX^\star = \bL^{-\dagger} \bPsi \bSig^\star $

\ifCLASSOPTIONcaptionsoff
  \newpage
\fi

\addcontentsline{toc}{chapter}{Bibliography}
\bibliography{ref_hadi_merged}
\bibliographystyle{ieeetr}

\end{document}